\documentclass[12pt]{article}
\usepackage{amssymb,graphicx,amsmath,array,verbatim,hyperref,
here}
\usepackage{soul}

\newcommand{\beq}{\begin{eqnarray}}
\newcommand{\eeq}{\end{eqnarray}}
\newcommand{\p}{\partial}
\newcommand{\NF}{N_{\rm F}}

\newcommand{\vs}[1]{\vspace{#1 mm}}
\newcommand{\hs}[1]{\hspace{#1 mm}}
\newcommand{\bpm}{\begin{pmatrix}}
\newcommand{\epm}{\end{pmatrix}}
\newcommand{\Z}{\mathbb{Z}}
\newcommand{\R}{\mathbb{R}}

\newcommand{\D}{\mathcal D}

\newcommand{\ba}{\left(\begin{array}}
\newcommand{\ea}{\end{array} \right)}

\renewcommand{\thefootnote}{\arabic{footnote}}

\makeatletter
\@addtoreset{equation}{section}

\makeatother

\usepackage[usenames]{color}

\begin{document}
\thispagestyle{empty}

\setstcolor{red}

\vspace*{-2cm}
\begin{flushright}
{\tt YGHP-18-07}
\end{flushright}

\begin{center}

\vspace{-.5cm}

\vspace{0.5cm}
{\bf\Large Domain Wall and Three Dimensional Duality}

\vspace{0.5cm}
Minoru Eto${}^{1}$,
Toshiaki Fujimori${}^{2}$ and
Muneto Nitta${}^{2}$

\vspace{0.5cm}
{\it\small 
${}^1$ Department of Physics, Yamagata University, Kojirakawa-machi 1-4-12, Yamagata, Yamagata 990-8560, Japan}\\
{\it\small
${}^2$ Department of Physics, and Research and Education 
Center for Natural Sciences,}\\  
{\it\small 
 Keio University, Hiyoshi 4-1-1, Yokohama, Kanagawa 223-8521, Japan}

\vspace{1cm}
{\bf Abstract} 
\vspace{1cm}

\begin{minipage}[h]{0.9\hsize}
We discuss 1/2 BPS domain walls 
in the 3d $\mathcal N=4$ supersymmetric gauge theory 
which is self-dual under the 3d mirror symmetry. 
We find that if a BF-type coupling is introduced, 
invariance of the BPS domain wall 
under the duality transformation
can be explicitly seen from the classical BPS equations.
It has been known that particles and vortices are 
swapped under the 3d duality transformations. 
We show that Noether charges and 
 vortex topological charges localized on the domain walls are 
correctly exchanged under the 3d mirror symmetry. 
\end{minipage}
\end{center}

\vspace{0.5cm}
\parbox{15cm}{
\small\hspace{15pt}
}


\clearpage
\setcounter{page}{1}
\setcounter{footnote}{0}
\renewcommand{\thefootnote}{\arabic{footnote}}
\newpage

\tableofcontents

\section{Introduction}
Three dimensional dualities are 
useful tools to study various aspects of quantum field theories 
in both high energy and condensed matter physics. 
It has been known that there exist duality transformations 
under which particles and topological vortices are exchanged \cite{Peskin:1977kp,Dasgupta:1981zz},
see Refs.~\cite{Son:2015xqa,Karch:2016sxi,Murugan:2016zal,Seiberg:2016gmd,Benini:2017aed} for recent developments.
Photons and scalar fields
which mediate long-range forces 
between charged particles and vortices are 
also exchanged under those duality transformations. 
The 3d mirror symmetry in supersymmetric models 
\cite{Intriligator:1996ex}, 
is an example of such particle-vortex dualities. 
It swaps a Coulomb branch of a supersymmetric gauge theory 
and a Higgs branch of the dual model. 
If those vacuum moduli spaces are lifted in such a way that 
only some discrete points remain supersymmetric vacua, 
there should be BPS domain wall solutions in both branches 
\cite{Abraham:1992vb,Abraham:1992qv,
Gauntlett:2000ib,Tong:2002hi,Shifman:2002jm,
Tong:2003ik,Isozumi:2003rp,Eto:2009wq,Tong:2005un,Eto:2006pg,Shifman:2007ce}.
Some properties of the domain walls 
under the duality transformation has been discussed and 
an interesting relation to the 2d mirror symmetry 
was pointed out \cite{Tong:2003ik}. 

In this paper, we discuss the duality property of 1/2 BPS domain walls 
from the viewpoint of classical BPS equations 
in 3d $\mathcal N = 4$ Abelian gauge theories. 
In a self-dual model such as 
SQED with $\NF=2$ charged hypermultiplets, 
domain walls are expected to be invariant 
under the duality transformation. 
However, their profiles look different 
when parameters of the model are transformed by the duality map.
One may think that the duality is valid only in the IR regime 
and it cannot be seen in the classical BPS configurations. 
However, it has been known that 
the duality can be seen at any energy scale 
if the model is modified by introducing a BF-type coupling \cite{Kapustin:1999ha}.
We study domain wall configurations 
in the modified models and 
compare them to see how domain wall profiles 
transform under the duality. 
Although BPS domain wall equations are not invariant 
under the duality map of the parameters, 
the duality is correctly reflected 
in the internal structure of domain wall 
which can be seen in classical configurations of the modified models. 

The organization of this paper is as follows. 
In Sec.\,\ref{sec:parton}, 
we review the BPS domain wall configuration 
in SQED with $\NF = 2$ hypermultiplets, 
which is known as a self-dual model. 
In Sec.\,\ref{sec:SD_model}, 
we modify the model by introducing a BF-type coupling 
and find that the duality is correctly reflected 
in classical domain wall configurations. 
In Sec.\,\ref{sec:QJ}, 
BPS domain wall configurations 
with Noether and vortex charges are discussed. 
We show that they are distributed  
on the domain wall in such a way that 
they are correctly exchanged under the duality transformation. 
Sec.\,\ref{sec:summary} is devoted to a summary and discussions. 

\section{1/2 BPS Domain Wall in 3d $\mathcal N = 4$ SQED}
\label{sec:parton}
\begin{figure}[h]
\begin{center}
\begin{minipage}[h]{0.45\hsize}
\begin{center}
\includegraphics[width=82mm]{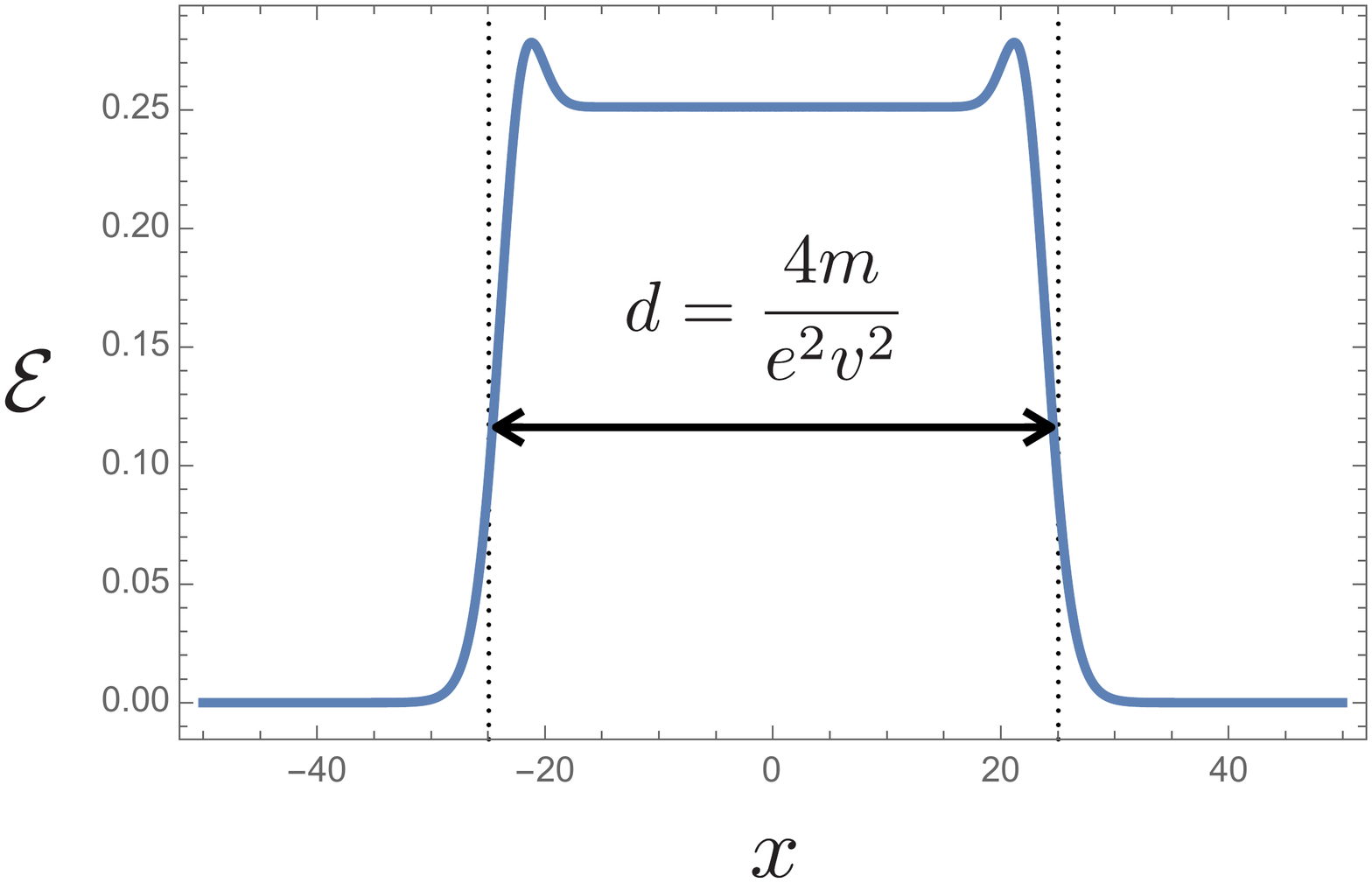} \\ \vs{-5}
(a) energy density $\mathcal E$
\end{center}
\end{minipage}
\hs{5}
\begin{minipage}[h]{0.45\hsize}
\begin{center}
\includegraphics[width=80mm]{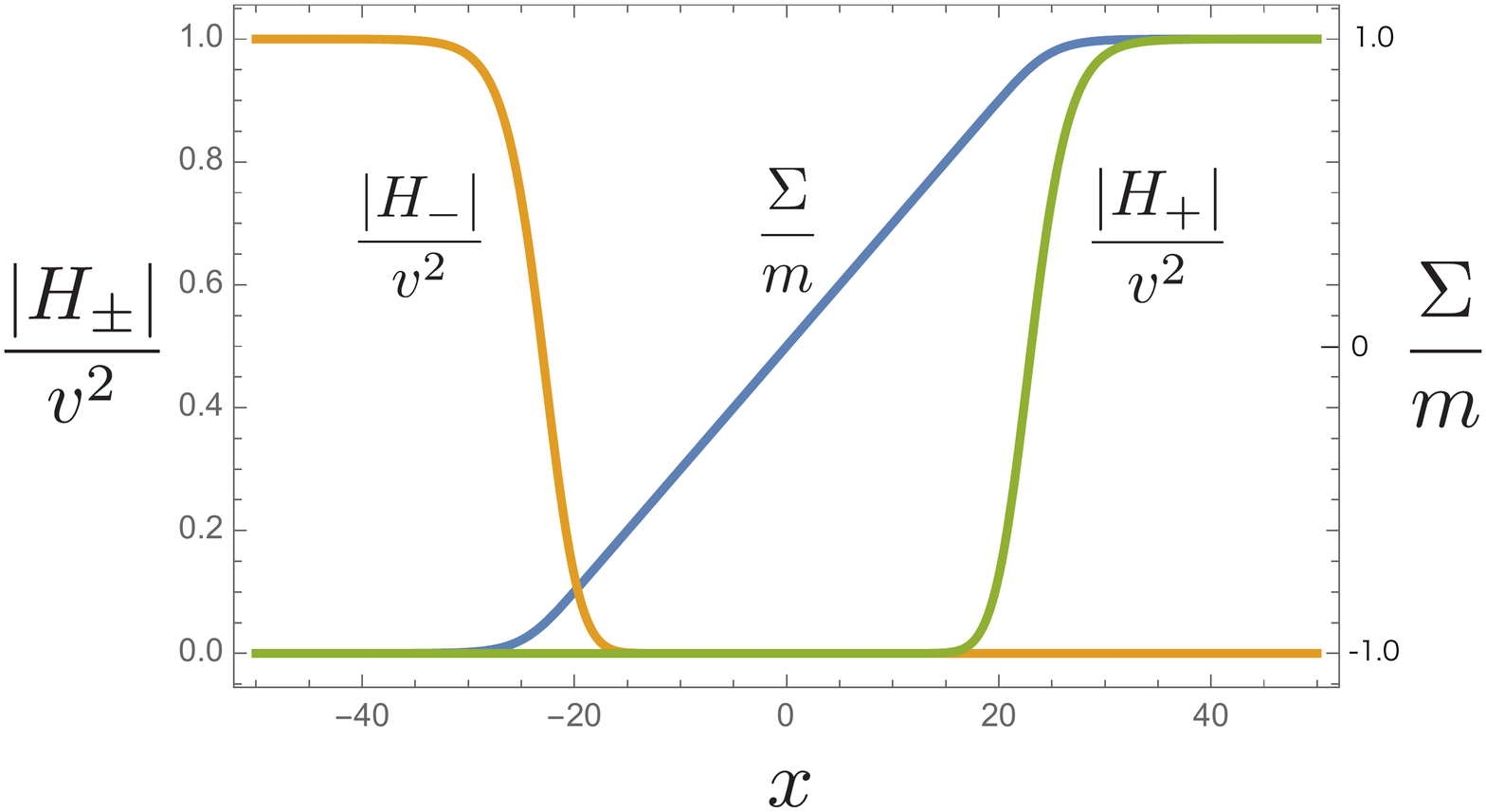} \\ \vs{-5}
(b) scalar fields $\Sigma$ and $|H_\pm|$
\end{center}
\end{minipage}
\caption{The profile of domain wall for $d=50$, $m=\pi$, $v^2=2$: 
(a) the energy density $\mathcal E = \p_x \left[ v^2 \Sigma - (\Sigma - m) |H_+|^2 - (\Sigma + m) |H_-|^2 \right]$ and 
(b) the scalar fields $\Sigma$ and $|H_\pm|$. }
\label{fig:wall_profile}
\end{center}
\end{figure}

In this section, we briefly recapitulate the 1/2 BPS domain wall 
in 3d $\mathcal N = 4$ SQED. 
For simplicity, we restrict ourselves 
to the simplest example of $U(1)$ gauge theory 
with two charged hypermultiplets (SQED with $\NF=2$),
where the BPS equations are given by 
\cite{Tong:2002hi}
\beq
\p_x H_+ = - (\Sigma - m) H_+, \hs{5} 
\p_x H_- = - (\Sigma + m) H_-, \hs{5} 
\p_x \Sigma = \frac{e^2}{2} (|H_+|^2 + |H_-|^2 - v^2) ,
\eeq
where $H_\pm$ and $\Sigma$ are 
the scalar components of the charged hypermultiplets 
and the vector multiplet, respectively. 
We have chosen the gauge fixing condition 
such that the gauge field in the $x$-direction vanishes $(A_x=0)$.
There are three parameters in this system: 
the gauge coupling constant $e$, 
the hypermultiplet mass $m$ 
and the Fayet-Iliopoulos (FI) parameter $v^2$. 
These equations have a domain wall solution 
interpolating the two degenerate vacua 
\beq
\big(\,\Sigma,\,H_+,\,H_- \big) = \big(m,\,v,\,0 \big) \hs{10} 
\mbox{and} \hs{10}
\big(\,\Sigma,\,H_+,\,H_- \big) = \big(-m,\,0,\,v \big).
\eeq
A domain wall profile in the weak gauge coupling regime 
($e^2 \approx 0$) is shown in Fig.\,\ref{fig:wall_profile}. 
In this regime, the energy density profile looks like 
a bound state of two constituents confined by an object 
with an uniform energy density (tension) 
\cite{Shifman:2002jm,Eto:2009wq}. 
They are stabilized at a finite distance, 
which can be estimated as follows.
In the weak coupling regime, 
the BPS kink solution can be approximated 
by the piecewise functions \cite{Shifman:2002jm}
\beq
(\Sigma, H_+, H_-) = \left\{ 
\begin{array}{cl} ( ~ m ~ , \ v \ , \ 0 \ ) & \hs{10} \mbox{for $x \ll 0$} \\ 
(\frac{2m}{d} x , \ 0 \ , \ 0 \ ) & \hs{10} \mbox{for $x \approx \, 0$} \\
(-m \ , \ 0 \ , \ v \ ) & \hs{10} \mbox{for $0 \ll x$} 
\end{array} \right., \hs{10}
d \equiv \frac{4m}{e^2 v^2},
\eeq
where we have fixed the center of mass position of the kink 
as $x_{\rm kink} = 0$. 
This approximate solution implies that 
the width of the wall, 
that is the distance between the two constituent objects, 
is given by the length scale parameter $d = \frac{4m}{e^2 v^2}$. 
Although it is unclear 
why such an internal structure appears
in the domain wall configuration of the current model, 
we will elucidate the origin of such a property of domain wall 
by making use of 3d mirror symmetry. 

\section{Domain Wall in the Self-Dual Model}
\label{sec:SD_model}

\subsection{Self-dual models}
Let us see what becomes of the domain wall 
under the 3d mirror symmetry transformation. 
Although the $U(1)$ gauge theory with $\NF=2$ 
is said to be self-dual, 
the domain wall width is not invariant 
under the mirror symmetry transformation, 
which swaps the FI parameter $v^2$ 
and the mass parameter $m$. 
This is because the self-duality of the current model 
is valid only in the IR limit. 
Therefore, to see the property of the domain wall 
under the mirror symmetry transformation, 
we have to modify the model 
so that the duality transformation is valid for all scale. 
In particular, we need to introduce 
a dual parameter for the coupling constant $g^2$. 

As discussed in \cite{Kapustin:1999ha}, 
such an extended self-dual theory can be obtained 
by coupling a twisted vector multiplet to 
two copies of $U(1)$ gauge theory 
with one charged hypermultiplet (SQED with $\NF=1$) 
via a BF-type coupling. 
By using the scalar-vector duality 
(see Appendix \ref{appendix:scalar_vector}),
the twisted vector multiplet can be rewritten into 
a hypermultiplet whose scalar components 
$(\chi, X, Y, Z)$ parametrize $S^1 \times \R^3$. 
Then the self-dual Lagrangian can be rewritten as
\beq
\mathcal L ~=~ \mathcal L^+_{\rm SQED} + \mathcal L^-_{\rm SQED} + \mathcal L_{\rm BF},
\label{eq:total_L}
\eeq
with
\beq
\mathcal L^\pm_{\rm SQED} \!&=&\! - \frac{1}{g_\pm^2} \left[ \frac{1}{2} (F_{\mu \nu}^\pm)^2 + (\p_\mu \Sigma_\pm)^2 + (D_\pm)^2 \right] - |\D_\mu H_\pm|^2 - (\Sigma_\pm - m_\pm)^2 |H_\pm|^2 + \cdots, \\
\mathcal L_{\rm BF} ~~ &=&\! - \ \frac{1}{2} \, \left[ u (\p_\mu X)^2 + \frac{1}{u} (\p_\mu \chi + A_\mu^+ - A_\mu^-)^2 + \frac{1}{u} (\Sigma_+ - \Sigma_-)^2 \right] + \cdots, 
\eeq
where $\cdots$ denotes terms 
which are irrelevant to domain wall solutions.  
The auxiliary fields $D_\pm$ are determined 
by solving the algebraic equations of motion as
\beq
D_{\pm} = \frac{g_{\pm}^2}{2} \left( |H_\pm|^2 \pm X - \xi_\pm \right). 
\eeq
Although the coupling constants $g_\pm$ can be different, 
in this paper, we set $g_+ = g_- = g$ for simplicity. 
Furthermore, shifting $\Sigma_\pm$ and $X$, 
we can always set
\beq
m_\pm = \pm m, \hs{10}
\xi_\pm = \xi. 
\eeq

The parameter $u$ corresponds to the radius of $S^1$ 
parametrized by the periodic scalar $\chi$ 
and it is related to the gauge coupling constant $\tilde e$ 
of the original twisted vector multiplet as $u \propto 1/\tilde e^2$. 
In the $u \rightarrow 0$ limit, 
this model reduces to the $N_F=2$ SQED discussed in the previous section. 
When $u=0$, we have to impose the following constraints 
so that $\mathcal L_{\rm BF}$ is finite
\beq
\p_\mu \chi + A_\mu^+ - A_\mu^-, =0 \hs{10} \Sigma_+ - \Sigma_- = 0.
\eeq
In addition, the kinetic term of $X$ disappears, 
i.e. $X$ becomes an auxiliary field. 
Integrating out $X$ and imposing 
the gauge fixing condition $\chi =0$, 
we can eliminate one of the vector multiplets. 
Thus, the resulting theory is identified with 
the $\NF=2$ SQED discussed in the previous section, 
where the parameters are related as
\beq
\frac{1}{e^2} = \frac{2}{g^2}, \hs{10} v^2 = 2 \xi.
\eeq

\paragraph{Coulomb and Higgs branches \\}
If either of the mass or FI parameters is sufficiently small, 
the low energy physics is described 
by the Coulomb or the Higgs branch effective theory 
with a shallow potential proportional to the small parameter.  
Both Coulomb and Higgs branch moduli spaces take 
the form of the two-center Taub-NUT space
whose asymptotic radius in the Coulomb and Higgs branches 
are respectively given by
\beq
R_{\rm Coulomb} = \frac{g^2}{4\pi^2}, \hs{20} R_{\rm Higgs} = \frac{1}{u}. 
\label{eq:S1_radius}
\eeq
The small FI and mass parameters give the following shallow potentials 
on the Coulomb and Higgs branches, respectively:
\beq
V_{\rm Coulomb} = \pi^2 \xi^2 ||\Xi||^2, \hs{10} V_{\rm Higgs} = m^2 ||\Xi||^2,
\label{eq:potential_moduli}
\eeq
where $||\Xi||^2$ denotes the squared norm of 
the tri-holomorphic Killing vector $\Xi$ 
on the two-center Taub-NUT space. 
It has been known that the two branches 
are swapped by the 3d mirror symmetry transformation
and the parameters are mapped as 
(see Appendix \ref{appendix:dual_pair} for details of the duality):
\beq
m \leftrightarrow \pi \xi, \hs{10} u \leftrightarrow \frac{4\pi^2}{g^2}.
\label{eq:duality}
\eeq 

\paragraph{Large and small $(g,u)$ limits \\} 
As we have seen above, our model reduces to 
the $U(1)$ gauge theory with two charged hypermultiplets 
(SQED with $\NF=2$) in the $u \rightarrow 0$ limit. 
The duality map Eq.\,\eqref{eq:duality} implies that 
the small $u$ limit corresponds to the large $g$ limit in the dual picture. 
In the $g \rightarrow \infty$ limit, 
both vector multiplets $(A_\mu^\pm, \Sigma_\pm, \cdots)$ become auxiliary fields 
and can be eliminated by solving their equations of motion.
The resulting effective model is the non-linear sigma model 
whose target space is the two-center Taub-NUT space (Higgs branch moduli space) 
with the potential proportional to $V_{\rm Higgs}$.

On the other hand, in the $u \rightarrow \infty$ limit, 
we have the constraint
\beq
\p_\mu X = 0,
\eeq
and the vector multiplets $(A_+, \Sigma_+,\cdots,)$ and $(A_-,\Sigma_-,\dots)$ are decoupled from each other. 
Therefore, the model becomes two copies of $U(1)$ gauge theories with a single charged hypermultiplets (two copies of SQED with $\NF=1$). 
The duality transformation \eqref{eq:duality} implies that 
this  limit corresponds to the small $g$ limit in the dual picture. 

In the following, we will see that domain walls 
in the large and small $(g,u)$ regimes have 
the identical properties as expected from the duality. 

\subsection{Domain wall solution}
\label{subsec:solution}
When both $\xi$ and $m$ are non-zero, 
the Lagrangian has two degenerate vacua, 
in which the VEVs of the scalar fields are given by
\begin{align}
\Sigma_\pm & = \phantom{-} m, &
|H_+|^2 &= 2 \xi, &
|H_-|^2 &= \, 0 \, , &
X &= - \xi, \\
\Sigma_\pm &= -m, &
|H_+|^2 &= \, 0 \, , &
|H_-|^2 &= 2 \xi, &
X &= \phantom{-} \xi.
\label{eq:vacua}
\end{align}
In this subsection, we discuss the property of the domain wall solutions
from the viewpoint of the duality. 

Let us first consider static domain wall configurations 
which depend only on a spacial coordinate $x$. 
The energy density for a static configuration 
can be rewritten into the Bogomol'nyi form
\beq
\mathcal E ~=~ \mathcal E_{\geq 0} + W_+ + W_-,
\eeq
where the positive semidefinite part $\mathcal E_{\geq 0}$ is given by
\beq
\mathcal E_{\geq 0} \,
= \, \sum_{i=+,-} \left[ \frac{1}{g^2} \left| \p_2 \Sigma_i + D_i \right|^2 
+ \left| \p_x H_i + (\Sigma_i - m_i) H_i \right|^2 \right] + \frac{u}{2} \left|\p_x X + u^{-1} (\Sigma_+ - \Sigma_-) \right|^2,
\eeq
with $m_\pm = \pm m$. 
The total derivative terms $W_\pm$, which correspond to the domain wall charges, are given by
\beq
W_\pm ~= - \p_x \left[ \frac{2}{g^2} \Sigma_\pm D_\pm \mp m | H_\pm|^2 \right]. 
\label{eq:wall_charge}
\eeq
Suppose that the field configurations at $x \rightarrow \pm \infty$ 
are given by the two different sets of the VEVs in Eq.\,\eqref{eq:vacua}. 
Then we find from the fact that $\mathcal E_{\geq 0}$ is positive semidefinite 
that the energy density satisfies  
\beq
\int dx \ \mathcal E ~\geq~ T ~\equiv \ \int dx \, (W_+ + W_-) ~=~ 4 m \xi. 
\label{eq:Tpm}
\eeq 
As expected, the tension $T$ is invariant under the duality map Eq.\,\eqref{eq:duality}.
This Bogomol'nyi bound is saturated if $\mathcal E_{\geq 0} = 0$, 
i.e. the following BPS equations are satisfied: 
\beq
\p_x \Sigma_\pm = - D_\pm, \hs{10}
\p_x H_\pm = - (\Sigma_\pm \mp m) H_\pm, \hs{10} 
\p_x X = - \frac{1}{u} (\Sigma_+ - \Sigma_-).
\label{eq:BPS}
\eeq
The last two equations can be solved  
by introducing profile functions $\psi^\pm$ as
\beq
\Sigma^\pm = \frac{1}{2} \p_x \psi_\pm, \hs{10} 
H^\pm = \sqrt{2\xi} \exp \left( \pm m x - \frac{1}{2} \psi^\pm \right), \hs{10}
X = - \frac{1}{2u} (\psi_+ - \psi_-). 
\eeq
The first BPS equations reduce 
to the following differential equations for the profile functions $(\psi_+, \psi_-)$: 
\beq
\p_x^2 \psi_\pm ~=~ g^2 \xi \left[ 1 - 2 e^{\pm 2 m x - \psi_\pm} \pm \frac{1}{2u\xi} (\psi_+ - \psi_-) \right].
\label{eq:master}
\eeq
The boundary conditions for $(\psi_+,\,\psi_-)$ have to be chosen 
so that \st{the} the solution \eqref{eq:BPS} approaches the vacua \eqref{eq:vacua} as $x \rightarrow \pm \infty$:
\beq
\psi_+ &\rightarrow& \phantom{-} 2 m x, \hs{10} 
\psi_- ~\rightarrow~ \phantom{-} \ 2 m x - 2 u \xi, \hs{10} 
\mbox{for $x \rightarrow +\infty$}, 
\label{eq:boundary1} \\
\psi_- &\rightarrow& - 2 m x, \hs{10}
\psi_+ ~\rightarrow~ - 2 m x - 2 u \xi , \hs{10} 
\mbox{for $x \rightarrow -\infty$}. 
\label{eq:boundary2}
\eeq 

\begin{figure}[t]
\begin{minipage}[c]{0.48\hsize}
\centering
\includegraphics[width=90mm]{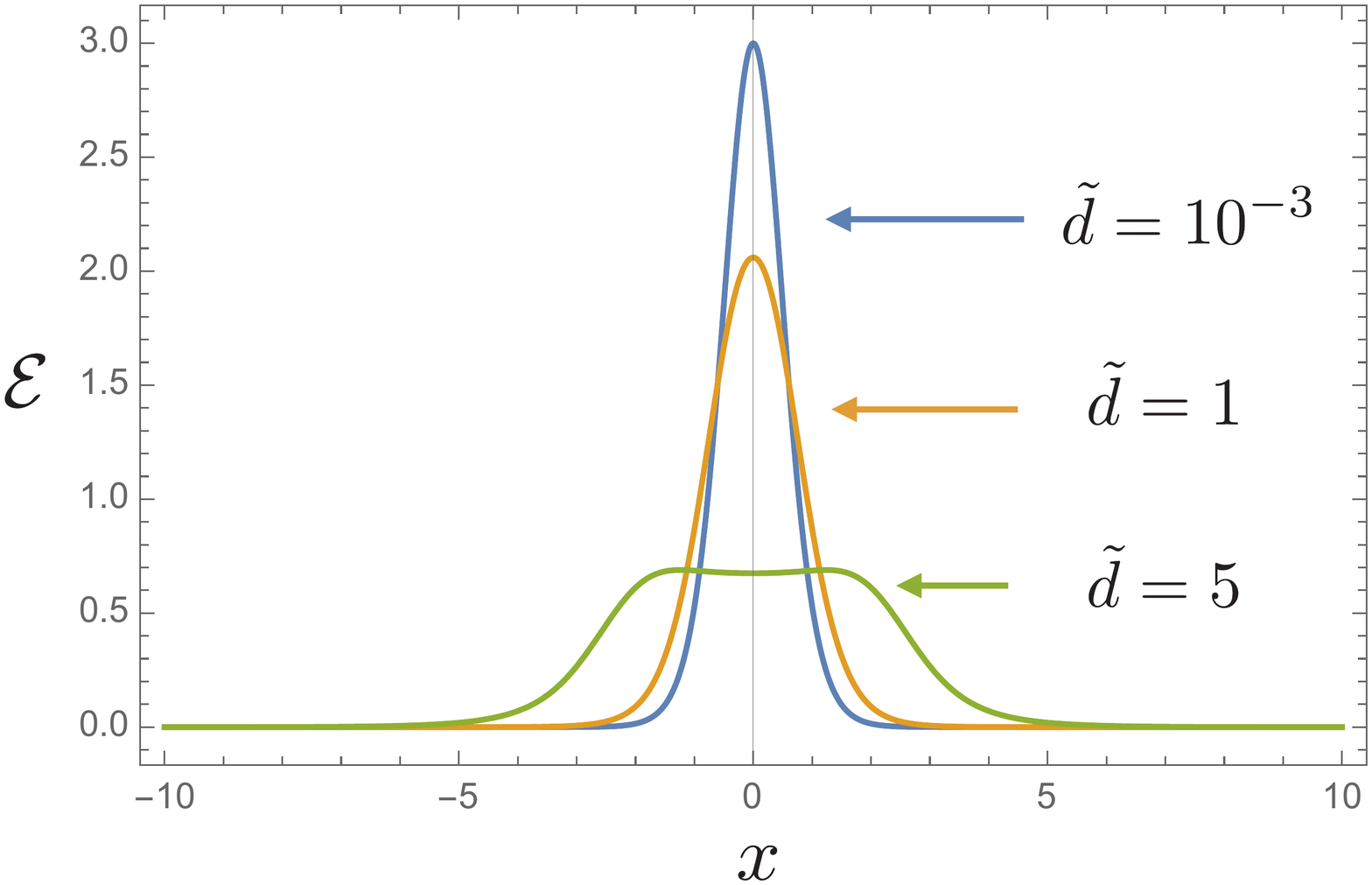} \\
~~ (a) $d = 1, ~ m^{-1} =1$
\end{minipage}
\begin{minipage}[c]{0.48\hsize}
\centering
\includegraphics[width=90mm]{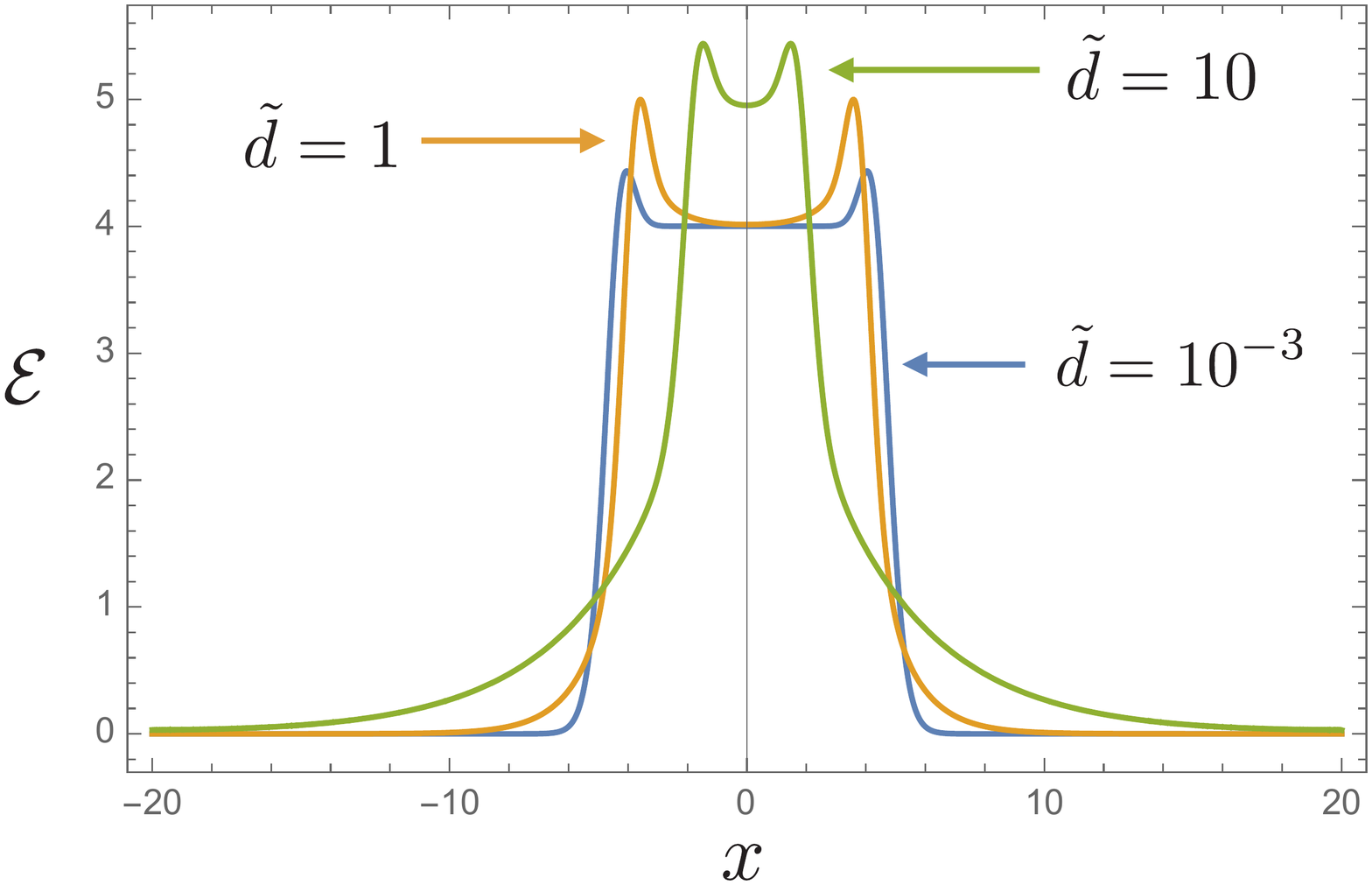} \\
~~~ (b) $d =10,~ m^{-1} =1$
\end{minipage}
\caption{Energy density profiles of domain wall configurations. 
For $d \lesssim m^{-1}$ (left), the wall width becomes larger as we increase $\tilde d$
and a plateau appears for sufficiently large $\tilde d$.
For $d \gg m^{-1}$ (right), 
the plateau region can be seen for small $\tilde d$ and 
it becomes smaller for larger $\tilde d$. 
}
\label{fig:energy}
\end{figure}
By introducing the dimensionless coordinate $y \equiv mx $, 
Eq.~\eqref{eq:master} can be rewritten as
\beq
\p_y^2 \psi_\pm ~=~ \frac{1}{md} \left[ 1 - 2 e^{\pm 2 y - \psi_\pm} \pm \frac{1}{2m \tilde d \, } (\psi_+ - \psi_-) \right], 
\label{eq:master2}
\eeq
where $d$ and $\tilde d$ are the characteristic length scales of the domain walls defined by
\beq
d = \frac{4m}{g^2 \xi}, \hs{10}
\tilde d = \frac{u \xi}{m}.
\eeq
Note that these two length \st{scales} scales are exchanged 
under the duality transformation \eqref{eq:duality}. 
The energy density of the BPS solution can be written in terms of 
the profile functions as
\beq
\mathcal E ~=~ \frac{m^2 \xi}{2} \p_y^2 \left[ \psi^+ + \psi^- - \frac{md}{4}  \p_y^2 \left( \psi_+ + \psi_- \right) + \frac{1}{4m \tilde d} (\psi_+-\psi_-)^2 \right].
\label{eq:energy_density}
\eeq

Figs.\,\ref{fig:energy}-(a),\,(b) shows 
the energy density profiles of the domain wall solutions 
for some typical values of the scale parameters. 
One of characteristic properties of these numerical solutions  
is that plateau regions appear in both 
large $(g,u)$ regime ($d \ll \tilde d$) and 
small $(g,u)$ regime ($\tilde d \ll d$). 

\paragraph{Width of domain wall \\}
We can see a self-duality of the domain wall 
from the widths of the plateau regions. 
As mentioned above, in the limit of small $u$ and $g$ $(\tilde d \ll m^{-1} \ll d)$, 
the profiles of $\Sigma_\pm = \frac{1}{2} \p_2 \psi_\pm$ becomes 
linear inside the domain wall $(x \approx 0)$. 
This can be seen from Eq.\,\eqref{eq:master}, 
which implies that the profile functions $\psi^\pm$ are 
approximately given by a quadratic function
\beq
\psi^+ \ \approx \ \psi^- \, \approx~ \frac{g^2 \xi}{2} x^2 + \cdots. 
\label{eq:approx_1}
\eeq
Since $\Sigma_\pm = m$ and $\Sigma_\pm = -m$ 
in the vacuum regions outside the domain wall, 
$\Sigma_\pm$ can be approximate as 
\beq
\Sigma_\pm ~\approx~ 
\left\{ 
\begin{array}{cl} 
- m & \mbox{left vacuum} \\ 
\frac{1}{2} g^2 \xi \, x & \mbox{inside wall} \\ 
~ m & \mbox{right vacuum} 
\end{array}
\right..
\eeq 
From the connectivity of the function $\Sigma$, 
the width of the wall $\Delta x$ can be estimated as\footnote{
It was known that the domain wall at weak gauge coupling regime in $u=0$ limit has the width $d$ \cite{Shifman:2002jm}.}
\beq
\frac{1}{2} g^2 \xi \Delta x = \Delta \Sigma_\pm = 2 m  ~~~ \Rightarrow ~~~ \Delta x = d. 
\eeq

On the other hand, when $u$ and $g$ are large ($d \ll m^{-1} \ll \tilde d$), 
the equation for the profile functions Eq.\,\eqref{eq:master2} implies that 
the scalar field $X \propto \psi_+ - \psi_-$ is a linear function inside the domain wall
\beq
\psi_+ - \psi_- ~\approx~ 4 m x + \cdots ~~~\, \Longrightarrow ~~~~ X ~\approx~ - \frac{2 m}{u} x + \cdots. 
\label{eq:approx_2}
\eeq
Since $X= -\xi$ and $X=\xi$ in the vacua, 
it can be approximated by the following piecewise linear function 
\beq
X ~\approx~ \left\{ 
\begin{array}{cl} 
~ \xi & \mbox{left vacuum} \\ 
- \frac{2 m}{u} x & \mbox{inside wall} \\ 
- \xi & \mbox{right vacuum} 
\end{array}
\right..
\eeq
From the connectivity of the function $X$, 
the width of the wall $\Delta x$ can be estimated as
\beq
- \frac{2 m}{u} \Delta x = \Delta X = -2 \xi  ~~~ \Rightarrow ~~~ \Delta x = \tilde d. 
\eeq

Therefore, the width of the domain wall is given 
by the length scale parameters $d$ and $\tilde d$ depending on the region in the parameter space: 
\beq
\Delta x = \left\{ 
\begin{array}{ll} 
\displaystyle d = \frac{4m}{g^2 \xi} \phantom{\Bigg(} & \mbox{for $\tilde d \ll m^{-1} \ll d$} \\ 
\displaystyle \tilde d= \ \frac{u \xi}{m} \phantom{\Bigg(} & \mbox{for $d \ll m^{-1} \ll \tilde d$} 
\end{array} \right..
\eeq
Since $d$ and $\tilde d$ are exchanged by the duality transformation \eqref{eq:duality}, 
the width of the domain wall is invariant under the duality. 

\begin{figure}[h]
\begin{center}
\includegraphics[width=80mm]{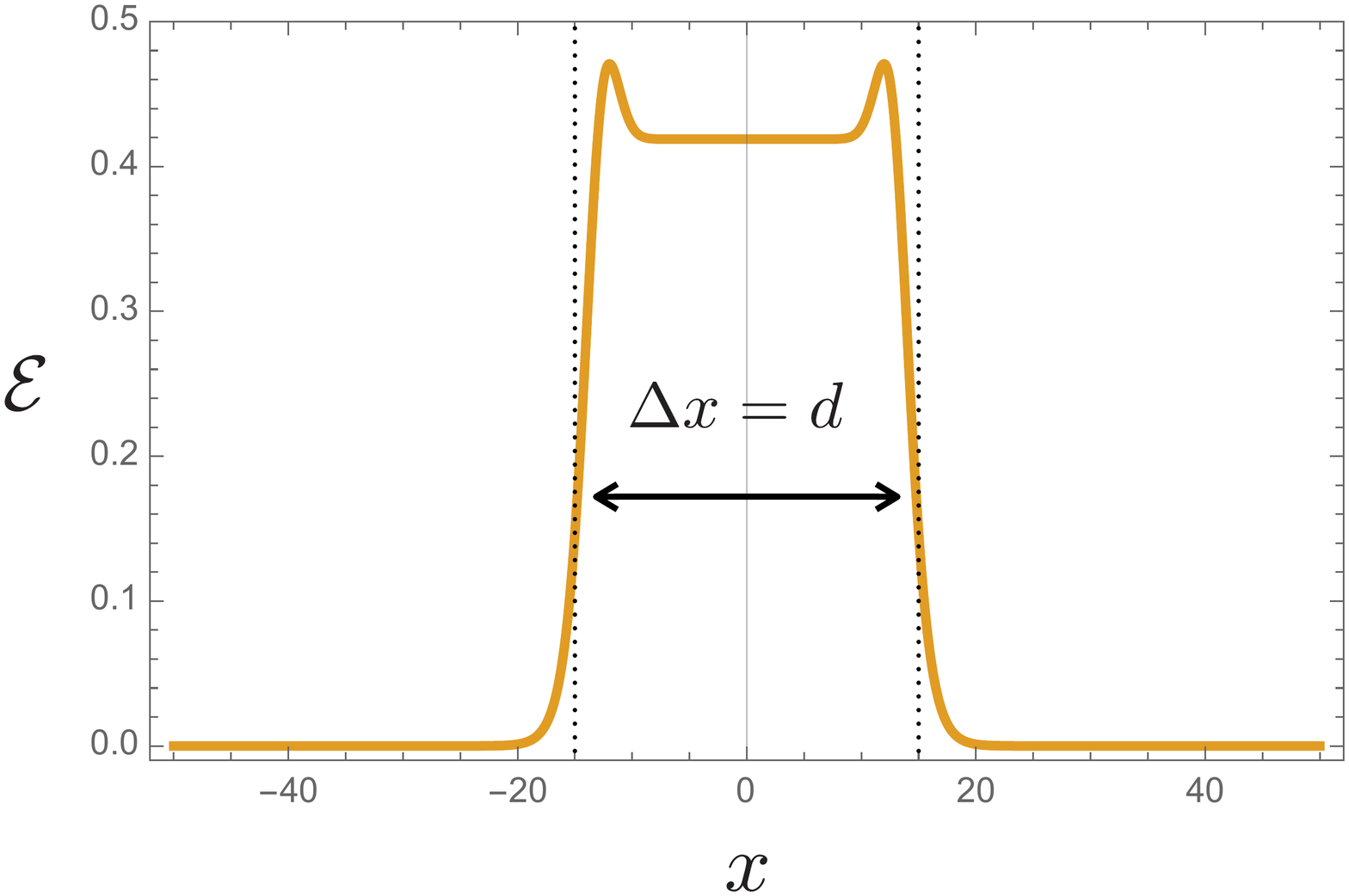} \hs{5}
\includegraphics[width=80mm]{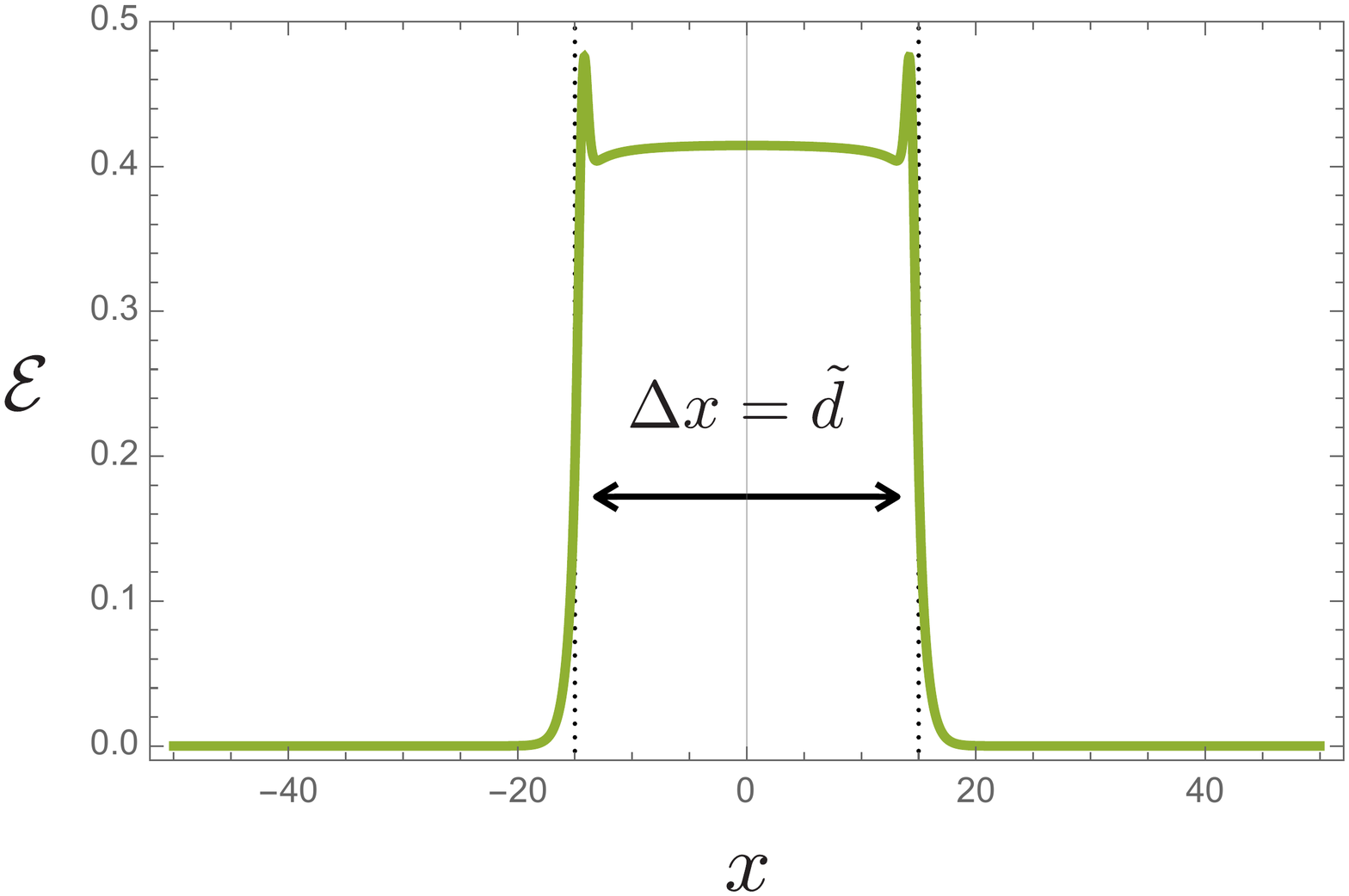} 
\caption{
The energy density profiles of the domain walls with 
$d=30,\,\tilde d=1/8$ (left) and 
$d=1/8,\,\tilde d=30$ (right). 
The mass and FI parameter are at the self-dual point $(m, \xi)=(\pi,1)$.}
\label{fig:width}
\end{center}
\end{figure}

Note that we find the self-dual property not only from the widths, 
but also from heights of the walls
(heights of the energy density at the plateau).
Plugging the approximate solutions $\psi_\pm$ 
into the energy density formula (\ref{eq:energy_density}), 
we find that the heights of the wall $h$ and $\tilde h$ 
for the parameter regions 
$\tilde d \ll m^{-1} \ll d$ and $d \ll m^{-1} \ll \tilde d$ 
are given by
\beq
h = g^2 \xi^2~~~(\mbox{for $\tilde d \ll m^{-1} \ll d$}), \hs{10} 
\tilde h = \frac{4m^2}{u}~~~(\mbox{for $d \ll m^{-1} \ll \tilde d$}). 
\eeq
As in the case of $d$ and $\tilde d$, 
$h$ and $\tilde h$ are also exchanged 
by the duality transformation \eqref{eq:duality},
so that the height of the wall is also invariant under the duality.
It is worth noting that the tension of the domain wall is also invariant 
since it can be written as $T = hd = \tilde h \tilde d$.
We show a typical example of the mirror pair of the small $(g,u)$ regime 
and of the large $(g,u)$ regime in Fig.\,\ref{fig:width}.

\begin{figure}[h]
\begin{center}
\includegraphics[width=17cm]{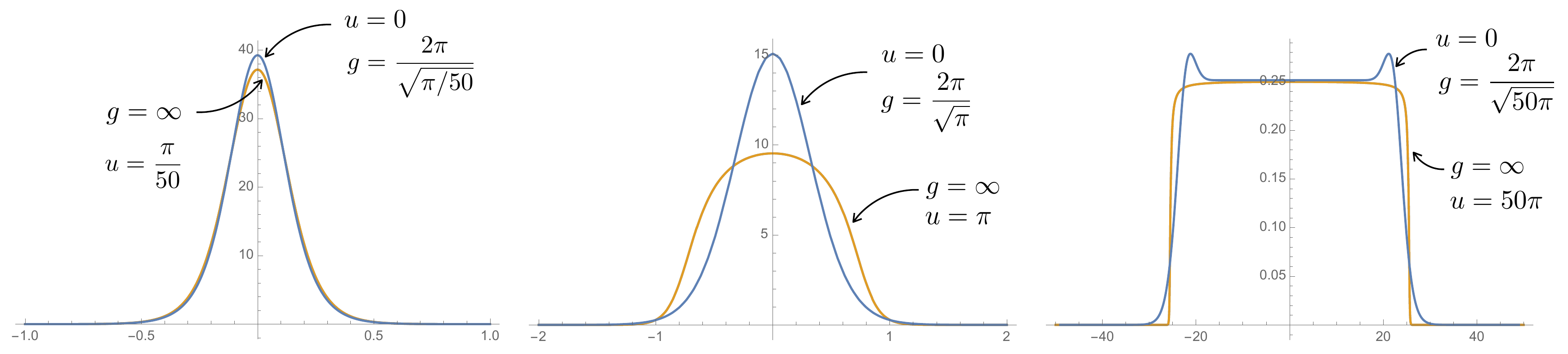}
\caption{
Energy density profiles of mirror pairs of domain walls (the blue lines for $u \to 0$ and the orange lines for $g \to \infty$). 
The mass and FI parameter are at the self-dual point $(\xi, m)=(1,\pi)$.}
\label{fig:walls}
\end{center}
\end{figure}

\paragraph{Duality between two-center Taub-NUT sigma model and $\NF=2$ SQED\\}
Although it is difficult to solve 
the coupled ordinary differential equations in Eq.\,(\ref{eq:master}), 
we can obtain analytic solutions in the strong gauge coupling limit 
by solving the following algebraic equation 
obtained from Eq.\,(\ref{eq:master}) in the $g \to \infty$ limit, 
\beq
1 - 2 e^{\pm 2 y - \psi_\pm} \pm \frac{1}{2m\tilde d} (\psi_+ - \psi_-) = 0.
\label{eq:master_g=infty}
\eeq
This equation describes the domain wall in the two-center Taub-NUT sigma model. 
The strong coupling limit corresponds to 
the $u \to 0$ limit in the dual picture, 
where the model reduces to SQED with $\NF=2$ hypermultiplets. 
In this case, $\psi_\pm$ must satisfy 
the constraint $\psi_+ = \psi_-$ and 
hence we are left with the ordinary differential equation
\beq
\psi = \psi_\pm, \quad\quad \p_y^2 \psi ~=~ \frac{4}{m d} \left[ 1 -  \left(e^{ 2 y} + e^{-2y}\right) e^{-\psi}\right] .
\label{eq:master_u=0}
\eeq
This equation is controlled by a dimensionless parameter $md$, 
and no analytic solutions has been found for generic $md$
except for several special discrete values \cite{Isozumi:2003rp}. 
Although Eq.\,(\ref{eq:master_g=infty}) is an algebraic equation and 
Eq.\,(\ref{eq:master_u=0}) is a differential equation,
the duality map (\ref{eq:duality}) implies that 
they describe essentially the same domain wall configuration.
We show some examples of 
dual pairs of domain walls in Fig.~\ref{fig:walls}.
One can see the widths of domain walls 
in the mirror pair are the same order
in the whole range of the parameters $(u,g)$. 

\begin{figure}[h]
\begin{center}
\includegraphics[width=80mm]{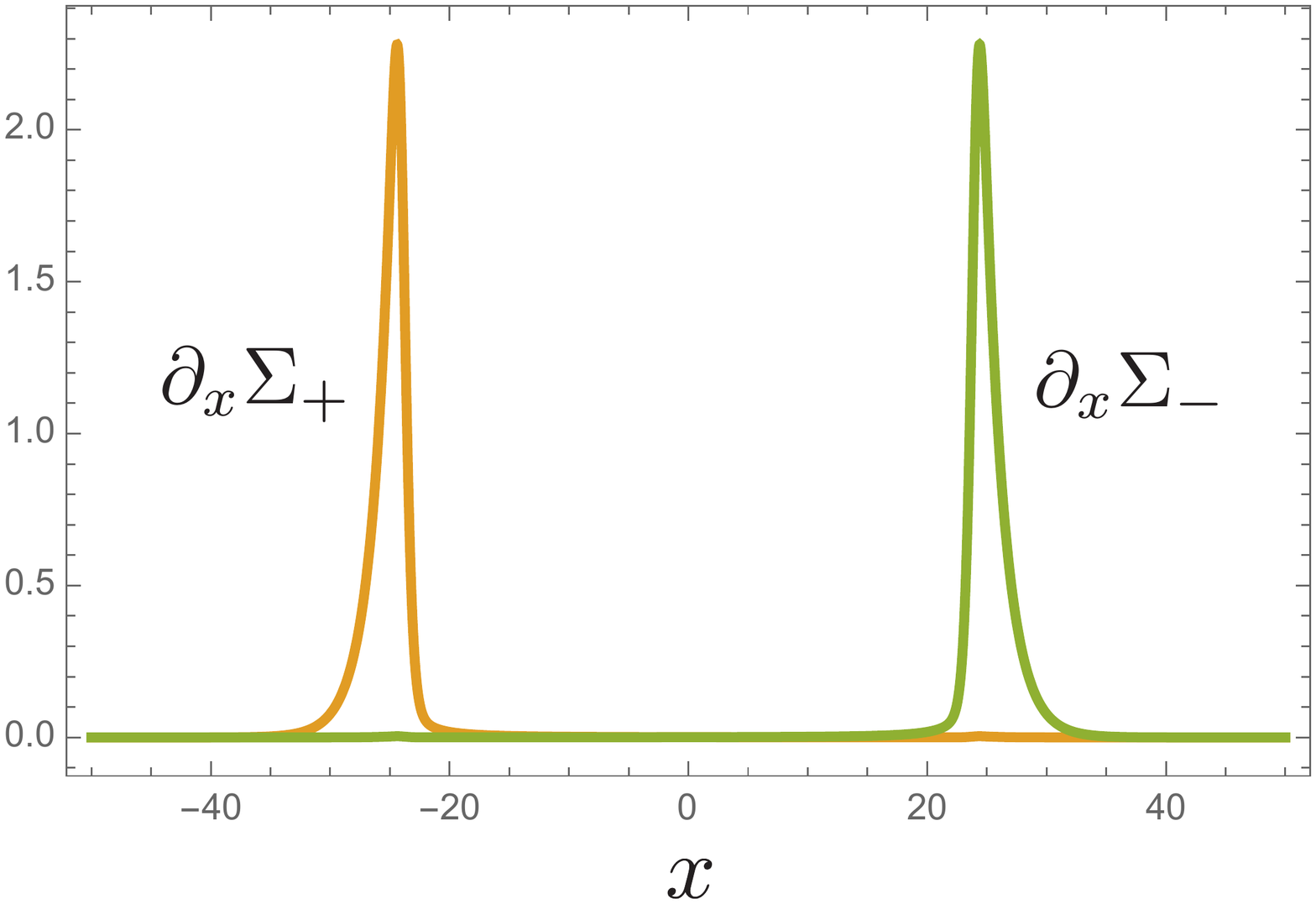} 
\includegraphics[width=80mm]{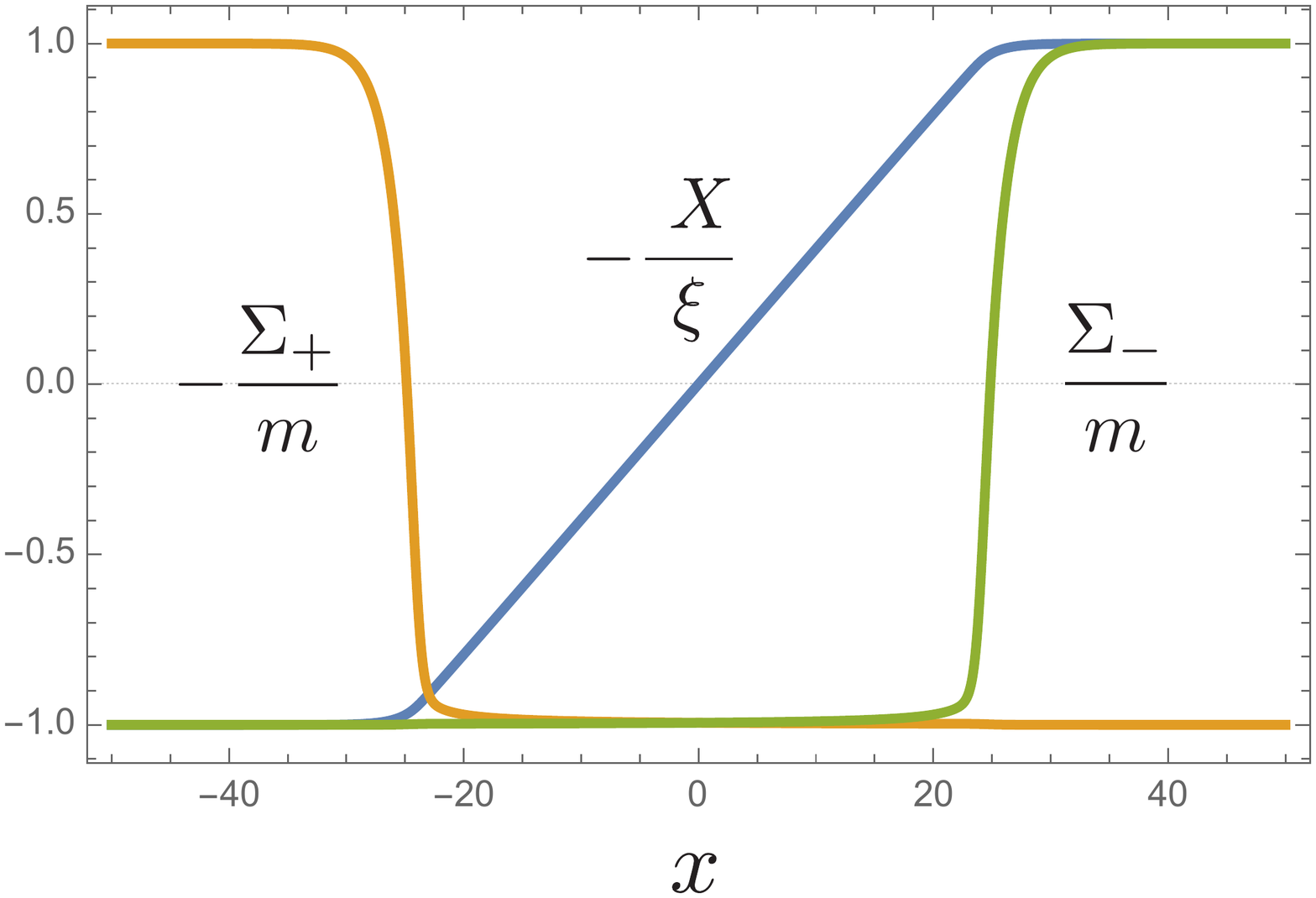} 
\caption{
The profiles of the kink topological charge density (left) and 
scalar fields (right) for $d=1/10,\,\tilde d=50$.
The mass and FI parameter are at the self-dual point $(m, \xi)=(\pi,1)$.}
\label{fig:fractional}
\end{center}
\end{figure}

\paragraph{The spikes in the energy density profiles \\}
In SQED with $\NF=2$ (the small $u$ limit), 
it has been known that 
there are spikes in the domain wall profile 
(see the left panel of Fig.\,\ref{fig:wall_profile} 
or the left panel of Fig.\,\ref{fig:width}). 
As expected from the duality, 
we can also see similar spikes in the dual picture 
(the right panel of Fig.\,\ref{fig:width}). 
Although the origin of such objects is unclear 
in the original picture ($\tilde d \ll m^{-1} \ll d$), 
we can identify them as a pair of confined domain walls
in the dual picture ($d \ll m^{-1} \ll \tilde d$). 
To see this, we first note that there are two types of walls
whose topological charges are given by Eq.\,\eqref{eq:wall_charge}
\beq
T_\pm ~=~ \int dx \, W_\pm ~=~ \xi \int dx \, \p_x \Sigma_\pm,
\eeq
where we have dropped some irrelevant terms in the integrand 
which do not contribute to $T_\pm$. 
As shown in Fig.\,\ref{fig:fractional}, 
in the dual picture, 
there are substructures of domain walls of $\Sigma_\pm$ in such a way that 
the topological charge densities 
are localized on the edges of the whole wall. 
Thus we can regard the whole domain wall 
as a bound state of the two constituent domain walls of $\Sigma_\pm$
confined due to the constant energy density of $X$ 
between them. 

Splitting of a single soliton 
to several partonic constituents is 
a common phenomenon which is frequently seen 
when it is deformed by taking a limit of parameters. 
A closely related model to ours is 3d lumps 
in the $N$-center Taub-NUT nonlinear sigma model. 
It was found that the single lump in the IR limit 
breaks up into $N$ partonic lumps 
with fractional topological charge $1/N$ \cite{Collie:2009iz}.
The kinks and lumps with fractional topological charges 
would be related to each other
in the same way as those with integer topological charges \cite{Eto:2006mz}.

\paragraph{Swapping of scalar fields \\}
It is worth noting that 
the profiles of the scalar fields for $\tilde d \ll m^{-1} \ll d$ 
(the right panel of Fig.\,\ref{fig:wall_profile})
and for $d \ll m^{-1} \ll \tilde d$ 
(the right panel of Fig.\,\ref{fig:fractional}) 
are almost identical if we identify the scalar fields as
\beq
|H_\pm| \leftrightarrow \mp \Sigma_\pm, \hs{10}
\Sigma_+ \approx \Sigma_- \leftrightarrow - X. 
\eeq 
This swapping of the scalar fields reflects the facts that   
chiral and vector multiplets are respectively mapped to 
(twisted) vector and chiral multiplets under the 3d mirror symmetry. 


\subsection{Effective actions and T-duality}

Next, let us consider the low energy effective theory 
on the domain wall. 
For later convenience, let $x_2$ be the transverse coordinate to the domain wall.
Since the translational symmetry 
$x_2 \rightarrow x_2 + x_0$ and 
the $U(1)$ global symmetry 
$H_\pm \rightarrow e^{\pm i \theta} H_{\pm}$
are broken by the domain wall, 
it has the position and phase moduli 
corresponding to the Nambu-Goldstone modes 
of the broken symmetries.
Therefore, the domain wall moduli space is a cylinder
\beq
\mathcal M = \R \times S^1,
\eeq
where $\R$ corresponds to the position $x_0$ and 
$S^1$ denotes the phase modulus $\theta$. 
In the thin wall limit, 
we can show that the domain wall worldsheet effective theory is described by the Nambu-Goto action
on the moduli space $\mathcal M$ \cite{Gauntlett:2000de,Eto:2015wsa}: 
\beq
\mathcal L_{\rm eff} ~=~ - T \sqrt{-\det g_{\alpha \beta}} ~=~ - 4 m \xi \sqrt{- \det \left( \eta_{\alpha \beta} + \p_\alpha x_0 \p_\beta x_0 + \frac{1}{m} \p_\alpha \theta \, \p_\beta \theta \right)},
\label{eq:NG1}
\eeq
where $\alpha$ and $\beta$ denote worldsheet indices. 
Let us consider the T-duality transformation along the $S^1$ direction. 
Writing $F_\alpha = \p_\alpha \theta$ and 
imposing the constraint $\epsilon^{\alpha \beta} \p_\alpha F_\beta =0$
by introducing a Lagrange multiplier $\tilde \theta$ as
\beq
\mathcal L_{\rm eff} = - T \sqrt{-\det g_{\alpha \beta}} + \frac{4}{\pi} \tilde \theta \, \epsilon^{\alpha \beta} \p_\alpha F_\beta,
\eeq
we can rewrite the effective Lagrangian by eliminating $F_\alpha$ as 
\beq
\mathcal L_{\rm eff} ~=~ - T \sqrt{-\det \tilde g_{\alpha \beta}} ~=~ - 4 m \xi \sqrt{- \det \left( \eta_{\alpha \beta} + \p_\alpha x_0 \p_\beta x_0 + \frac{1}{(\pi \xi)^2} \p_\alpha \tilde \theta \, \p_\beta \tilde \theta \right)},
\label{eq:NG2}
\eeq
where we have solved the equation of motion for $F_\alpha$
\beq
\frac{\delta S_{\rm eff}}{\delta F_\beta} ~=~ 0 ~~~~~ \Longrightarrow ~~~~~ 
F_\alpha = \p_\alpha \theta = \frac{m}{\pi \xi} \frac{\tilde g_{\alpha \beta} \epsilon^{\beta \gamma} \p_\gamma \tilde \theta}{\sqrt{-\det \tilde g_{\alpha \beta}}} .
\label{eq:F_sol}
\eeq
The T-dual pair of actions \eqref{eq:NG1} and \eqref{eq:NG2} are related 
by the swapping of the parameter $m \leftrightarrow \pi \xi$,
which ensures that 
the domain wall worldsheet theory is 
invariant under the 3d mirror symmetry. 

Both the original effective theory \eqref{eq:NG1} 
and the dual effective theory \eqref{eq:NG2}
have BPS solutions 
\beq
\theta = \omega t + k x, \hs{10}
\tilde \theta = \tilde \omega t + \tilde k x,\qquad (x \equiv x_1),
\eeq
where $\p_\alpha x_0 = 0$ and 
$(\omega, \, k)$ and $(\tilde \omega, \, \tilde k)$ are constants
corresponding to the internal momentum and the winding number. 
They are dual to each other 
if $(\omega, k)$ and $(\tilde \omega, \tilde k)$ satisfy 
the following relation so that Eq.\,\eqref{eq:F_sol} is satisfied
\beq
(\omega, k) = - \frac{m}{\sqrt{(\pi \xi)^2 - \tilde \omega^2 + \tilde k^2}} (\tilde k , \tilde \omega). 
\eeq
From this relation, we can show the agreement of the tension 
of these BPS states 
\beq
T_{\omega, k} 
~=~ 4 \xi \frac{m^2 + k^2}{\sqrt{m^2 - \omega^2 + k^2}} 
~=~ \frac{4m}{\pi} \frac{(\pi \xi)^2 + \tilde k^2}{\sqrt{(\pi \xi)^2 - \tilde \omega^2 + \tilde k^2}}.
\label{eq:tension_QJ_eff}
\eeq
This swapping of the internal momentum and the winding number 
can be regarded as an exchange of charges of the domain wall
from the balk viewpoint. 
In the next section, 
we discuss the duality property of 
such excited domain wall configurations. 

\section{Domain walls with Noether and vortex charges}
\label{sec:QJ}
In the previous section, 
we have seen that the internal momentum and 
the winding number of the excited domain wall states
are exchanged by the duality transformation. 
From the bulk viewpoint, they correspond to 
the Noether charge of the global $U(1)$ symmetry 
\cite{Abraham:1992vb,Abraham:1992qv}
and the vortex topological charge associated 
with the broken $U(1)$ gauge symmetry \cite{Eto:2015vsa}.
As mentioned above, 
it is well-known that such Noether and topological charges are 
exchanged under the duality transformation (particle-vortex duality).
In this section, we discuss the duality property of the domain wall with Noether and vortex charges. 

Let us consider stationary domain wall configurations 
characterized by the internal phase frequency 
and wave number $(\omega,k)$. 
In this section, 
$x_\mu~(\mu=0,1)$ and $x_2$ denote 
the coordinates along the domain wall worldsheet 
and the codimension, respectively. 
For later convenience, 
let us define a parameter $M$ by
\beq
M \equiv \sqrt{m^2 - \omega^2 + k^2}. 
\eeq 
Suppose that the Gauss law equations are satisfied
\beq
0 &=& \frac{2}{g^2} \p_i F_{0 i}^\pm + i ( H_\pm \D_0 \bar H_\pm - \bar H_\pm \D_0 H_\pm ) 
\pm \frac{1}{u} \left( \p_0 \chi + A_0^+ - A_0^- \right) . \phantom{\Bigg]}
\eeq
Then the energy density of the system can be decomposed into
\beq
\mathcal E &=& \mathcal E_0 + \mathcal E_+ + \mathcal E_- + \, \mathcal T_{\omega,k} \, + \, \{\mbox{total derivative} \}, 
\eeq
where $\mathcal T_{\omega,k}$ is the following combination of topological charges and Noether charges 
\beq
\mathcal T_{\omega,k} = \frac{m^2-\omega^2}{mM} (W_+ +W_-) + \frac{k}{M}(V_+^0 + V_-^0) + \frac{\omega}{M} (J_0^+ - J_0^-).
\eeq
This quantity gives the lower bound of the energy 
$\int dx_2 \, \mathcal E \geq \int dx_2 \, \mathcal T_{w,k}$
determined by the domain wall charges $W_\pm$ 
in \eqref{eq:wall_charge}
and $(V_\pm^0, J^{\pm}_0)$ are zeroth components of 
the vortex topological current $V_\pm^\mu$ 
are the Noether currents $J_\mu^\pm$ 
associated with the phase rotations of the scalar fields $H_\pm$,
\beq
V_\pm^\mu ~= \epsilon^{\mu \nu \rho} \p_\nu \left( \xi A_\rho^\pm - i H_\pm \D_\rho \bar H_\pm \right), \hs{10} 
J_\mu^\pm ~=~ i M ( H_\pm \D_\mu \bar H_\pm - \bar H_\pm \D_\mu H_\pm ).
\eeq
The total derivative terms are given by
\beq
\{\mbox{total derivative} \} = \frac{\omega}{m} \p^i \left( \frac{1}{g_+^2} \Sigma_+ F^+_{0 i} + \frac{1}{g_-^2} \Sigma^- F^-_{0 i} \right) + \frac{k}{M} \epsilon^{ij} \p_i \left( X \D_j \chi \right),
\eeq
where we have defined 
$\D_\mu \chi \equiv \p_\mu \chi + A_\mu^+ - A_\mu^-$. 
The positive semi-definite terms 
$\mathcal E_e$, $\mathcal E_+$ and $\mathcal E_-$ 
(see Appendix \ref{appendix:BPSeq})
vanish when the following BPS equations are satisfied 
\beq
F_{02}^\pm = - \frac{\omega}{M} D_\pm, \hs{10}
\D_0 H_\pm \hs{-2} &=& \hs{-2} - i \frac{\omega}{m} (\Sigma_\pm \mp m) H_\pm, \hs{10}
\D_0 \chi = - \frac{\omega}{m} (\Sigma_+-\Sigma_-), \phantom{\Bigg[} \\
F_{12}^\pm = - \frac{k}{M} D_\pm, \hs{10}
\D_1 H_\pm \hs{-2} &=& \hs{-2} - i \frac{k}{m} (\Sigma_\pm \mp m) H_\pm, \hs{10}
\D_1 \chi = - \frac{k}{m} (\Sigma_+ - \Sigma_-), \phantom{\Bigg[} \\
\p_2 \Sigma = - \frac{m}{M} D_\pm, \hs{10}
\D_2 H_\pm \hs{-2} &=& \hs{-2} - \frac{M}{m} (\Sigma_\pm \mp m) H_\pm, \hs{10} 
\p_2 X = - \frac{1}{u} \frac{M}{m} (\Sigma_+ - \Sigma_-). \phantom{\Bigg[}
\eeq
As in the case of the static domain wall, 
the BPS solution can be formally written as
\beq
\Sigma^\pm = \frac{1}{2} \frac{m}{M} \p_2 \psi_\pm, \hs{5} 
A_0^\pm = -\frac{1}{2} \frac{\omega}{M} \p_2 \psi_\pm, \hs{5} 
A_1^\pm = -\frac{1}{2} \frac{k}{M} \p_2 \psi_\pm, \hs{5} 
A_2^\pm = 0,
\eeq
\vs{-5}
\beq
H^\pm = \sqrt{2\xi} \exp \left[ \pm \left( i \omega t + i k x_1 + m x_2 \right) - \frac{1}{2} \psi_\pm \right], \hs{10}
X = - \frac{1}{2u} (\psi_+ - \psi_-),
\eeq
where $(\psi_+,\,\psi_-)$ are the functions satisfying  
\beq
\p_2^2 \psi_\pm ~=~ g_\pm^2 \xi \left[ 1 - 2 e^{\pm 2 M x_2 - \psi_\pm} \pm \frac{1}{2u\xi} (\psi_+ - \psi_-) \right].
\label{eq:masterQJ}
\eeq
These equations for the profile functions are the same as 
those for the static domain wall \eqref{eq:master} 
except that the mass $m$ is replaced by $M$. 
We can obtain profiles of domain wall configurations with 
Noether and vortex charges by solving Eq.\,\eqref{eq:masterQJ} 
with the analogous boundary conditions as the static case: 
\beq
\psi_+ &\rightarrow& \phantom{-} 2 M x_2, \hs{10} 
\psi_- ~\rightarrow~ \phantom{-} \ 2 M x_2 - 2 u \xi, \hs{10} 
\mbox{for $x_2 \rightarrow +\infty$}, 
\label{eq:boundaryQJ1} \\
\psi_- &\rightarrow& - 2 M x_2, \hs{10}
\psi_+ ~\rightarrow~ - 2 M x_2 - 2 u \xi , \hs{10} 
\mbox{for $x_2 \rightarrow -\infty$}. 
\label{eq:boundaryQJ2}
\eeq 

\begin{figure}[h]
\begin{center}
\includegraphics[width=16cm]{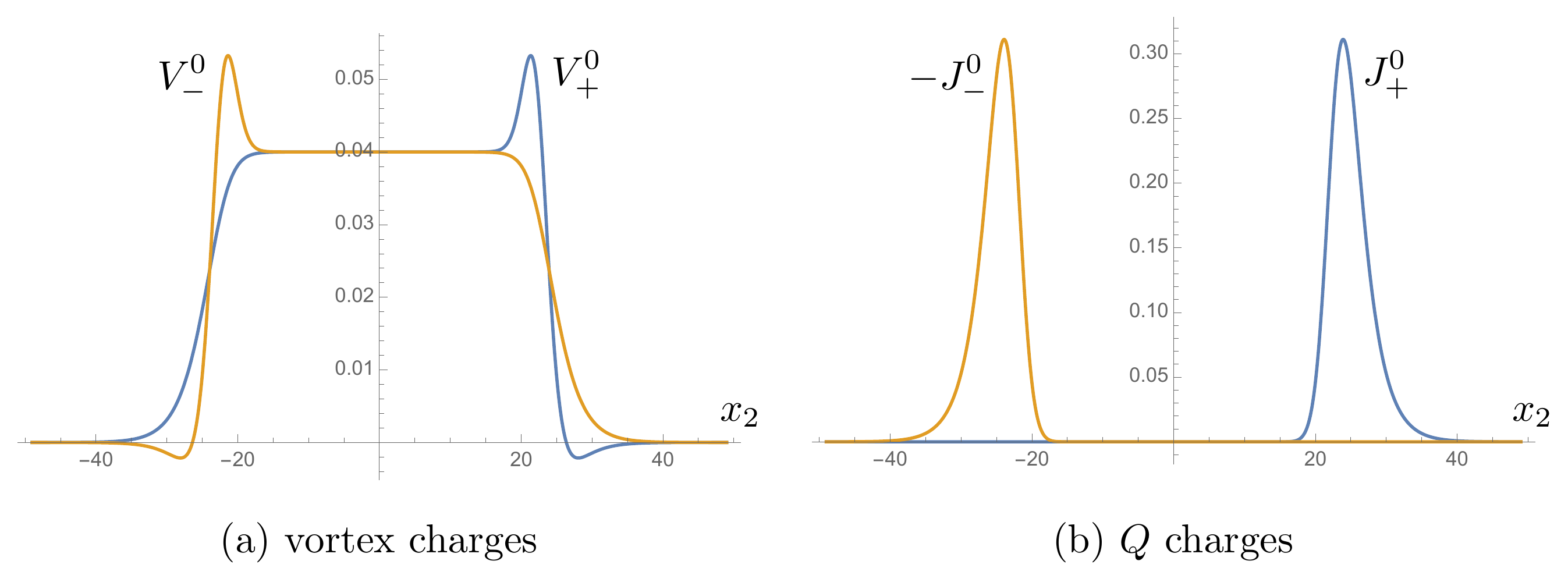}
\caption{Vortex and Noether charge densities with $d=50$, $\tilde d=1/10$, $m=\pi$, $\xi=1$, $(\omega,k)=(1,1)$.}
\ \\
\label{fig:d50dt01}
\includegraphics[width=16cm]{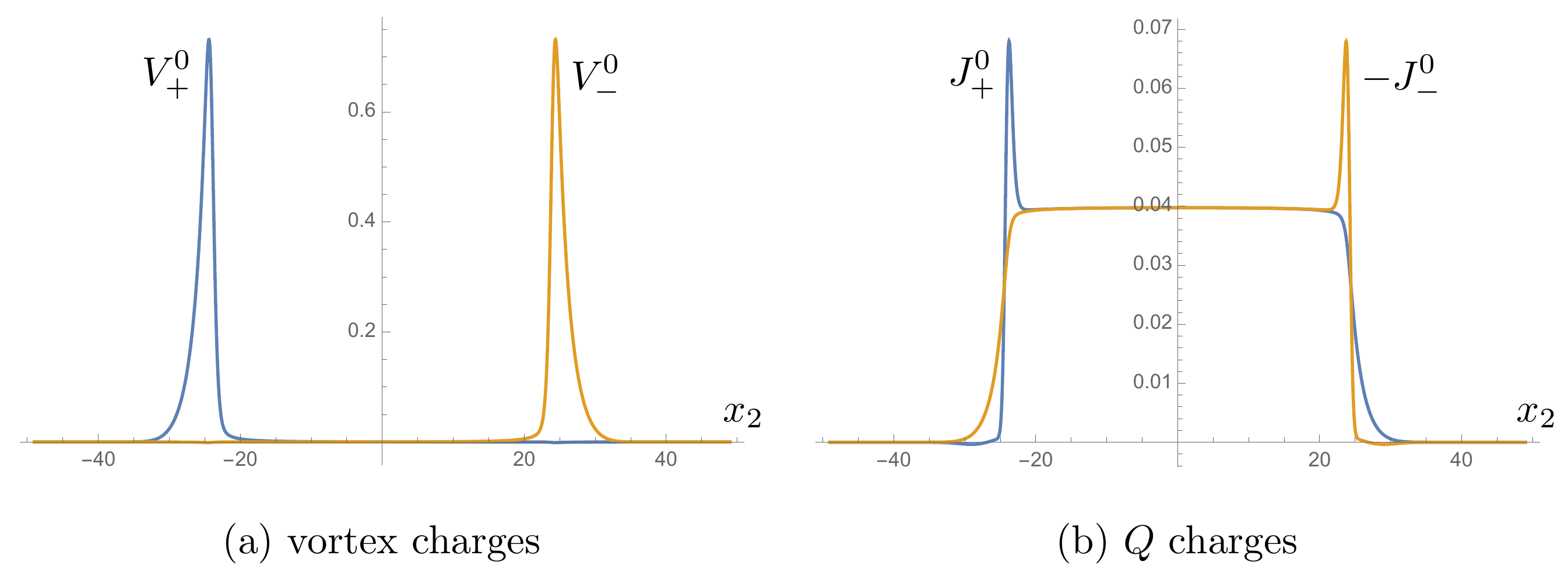}
\caption{Vortex and Noether charge densities with $d=1/10$, $\tilde d=50$, $m=\pi$, $\xi=1$, $(\omega,k)=(1,1)$.}
\label{fig:d01dt50}
\end{center}
\end{figure}

The 3d mirror symmetry implies that
the vortex topological currents $V_{\pm}^\mu$ and
the Noether currents $J_{\pm}^\mu$ are exchanged under the duality transformation. 
In terms of the profile functions, 
they are given by
\beq
V_\pm^0 &=& \frac{k}{2M} \p_{x_2}^2 \left[ - \frac{1}{g^2} \p_{x_2}^2 \psi_\pm \pm \frac{1}{2u} (\psi_+ - \psi_-) + \xi \psi_\pm \right], \\
J_0^\pm &=& \omega \p_{x_2} \left[ - \frac{1}{g^2} \p_{x_2}^2 \psi_\pm \pm \frac{1}{2u} (\psi_+ - \psi_-) \right].
\eeq
Since these quantities are total derivatives, 
we can integrate the charge densities 
by using the boundary conditions 
Eqs.\,\eqref{eq:boundaryQJ1} and \eqref{eq:boundaryQJ2} as
\beq
\int dx_2 \, V_{\pm}^0 = 2 \xi k, \hs{10} 
\int dx_2 \, J_0^\pm = \pm 2 \xi \omega. 
\eeq
Then we can check that the domain wall tension agrees 
with that of the BPS state in the effective theory 
in Eq.\,\eqref{eq:tension_QJ_eff}
\beq
T_{\omega, k} ~=~ \int dx \, \mathcal T_{\omega, k} ~=~ 4 \xi \frac{m^2 + k^2}{M}. 
\eeq

Since the equation for the profile function Eq.\,\eqref{eq:masterQJ}
is essentially the same as the corresponding equation 
in the static case Eq.\,\eqref{eq:master}, 
we can obtain approximate solutions for the wall 
with Noether and vortex charges 
from those for the static domain wall 
Eq.\,\eqref{eq:approx_1} and Eq.\,\eqref{eq:approx_2}
by replacing $m$ with $M$ . 
For $d \gg m^{-1} \gg \tilde d$,  
the vortex charge densities are constant inside the domain wall 
and the Noether charge densities are localized on the edges of the wall
as shown in the numerical solution in Fig.\,\ref{fig:d50dt01}. 
On the other hand, for $\tilde d \gg m^{-1} \gg d$, 
they are localized in the opposite way: 
the vortex charge densities are concentrated on the edges and 
Noether charge densities spread out inside the wall 
as shown in Fig.\,\ref{fig:d01dt50}.
Comparing the numerical solutions 
Figs.\,\ref{fig:d50dt01} and \ref{fig:d01dt50}, 
one sees that, as expected, 
$V_\pm^0$ and $\mp J_\mp^0$ are swapped under the duality.
Furthermore, We can analytically show that the height of 
the vortex charge densities $h_V$ and
the Noether charge densities $h_J$ are given by
\beq
h_V = \left\{ 
\begin{array}{cc} 
\frac{k}{2M} g^2 \xi^2 & \mbox{for $d \gg m^{-1} \gg \tilde d$} \\
0 & \mbox{for $\tilde d \gg m^{-1} \gg d$} 
\end{array} \right., \hs{10}
h_J = \left\{ 
\begin{array}{cc} 
0 & \mbox{for $d \gg m^{-1} \gg \tilde d$} \\
\frac{2M \omega}{u}  & \mbox{for $\tilde d \gg m^{-1} \gg d$} 
\end{array} \right.,
\eeq
and these quantities consistently transform 
under the duality transformation. 
Thus, we can check the duality 
by looking at the localization properties of 
the vortex charges and the Noether charges on the BPS domain wall. 


\section{Summary and Discussion}
\label{sec:summary}
In this paper, we have discussed the 1/2 BPS domain wall 
in the 3d $\mathcal N=4$ supersymmetric gauge theory 
which is self-dual under the 3d mirror symmetry. 
We have checked the BPS domain wall is self-dual 
and shown that the width, height, shape and tension of the wall are 
invariant under the duality transformation. 
We have shown that the domain wall 
in $\NF=2$ SQED (small $u$ limit) can be seen 
as a pair of confined fractional domain walls 
in the dual two-center Taub-NUT sigma model. 
We have seen that as expected from the vortex-particle duality,
the Noether charges and the vortex topological charges 
are correctly exchanged under the 3d mirror symmetry. 

We can also generalize the discussion 
to models with more Abelian gauge fields and matters. 
In such a case, the dual model is a different system. 
It would be interesting to see
how domain walls in different systems are related to each other
and discuss the connection between 3d and 2d mirror symmetries
from the viewpoint of domain wall effective theories
as was done in SQED with $\NF$ flavors ($u= 0$) and 
multi-center Taub-NUT sigma model ($g=\infty$) \cite{Tong:2003ik}. 
Generalization to non-Abelian gauge groups such as $U(N)$ 
is one important direction, 
which may be doable 
since BPS domain walls in the Higgs branch of $U(N)$ gauge theories were studied 
\cite{Shifman:2003uh,Isozumi:2004jc,Isozumi:2004va,Eto:2004vy,Hanany:2005bq}.

Another interesting direction to be explored 
is the generalization to 1/4 BPS states such 
as domain wall webs \cite{Eto:2005cp,Eto:2005fm}.
It has been known in general that there are two types of 1/4 BPS configurations 
which preserve different combinations of the supercharges \cite{Eto:2005sw}: 
one preserving  $(1,1)$ supersymmetry, which we called type-IIa,
and the other preserving $(2,0)$ supersymmetry, which we called type-IIb,  
in the cases of 2d worldvolume. 
While the latter can be solved by the moduli matrix \cite{Eto:2006pg}, 
the former is difficult to solve \cite{Naganuma:2001pu} in the present stage. 
Since the 3d mirror symmetry exchanges these two combinations
it is expected that two types of 1/4 BPS configurations 
are swapped under the duality transformation. 
This may offer a tool to solve 1/4 BPS equations of type-IIa.
It would be also interesting to see 
how the 3d mirror symmetry plays 
a role in the effective theories of the domain wall web \cite{Eto:2006bb,Eto:2007uc}. 

\section*{Acknowledgement}
This work is supported by the Ministry of Education, Culture, Sports, 
Science (MEXT)-Supported 
Program for the Strategic Research Foundation at Private Universities ``Topological Science'' 
(Grant No. S1511006).
The work of M.~E. is supported by KAKENHI Grant Numbers 26800119, 
16H03984 and 
by a Grant-in-Aid for Scientific Research on Innovative Areas
“Discrete Geometric Analysis for Materials Design” No. JP17H06462 from 
the MEXT of Japan.
The work of M.~N. is also supported in part by 
JSPS Grant-in-Aid for Scientific Research (KAKENHI Grant No. 16H03984 
and 18H01217), 
and by a Grant-in-Aid for Scientific Research on Innovative Areas ``Topological Materials
Science'' (KAKENHI Grant No.~15H05855)  from the MEXT of Japan.

\appendix


\section{Scalar-Vector Duality}
\label{appendix:scalar_vector}
We can show that 
an Abelian gauge field and a periodic scalar field 
are dual to each other as follows. 
Consider a periodic scalar field $\gamma$ 
\beq
\mathcal L = - \frac{1}{2 u} \p_\mu \gamma \p^\mu \gamma. 
\label{eq:S_scalar}
\eeq
This Lagrangian can be obtained from
\beq
\mathcal L = - \frac{1}{2u} f_\mu f^\mu + \frac{1}{2\pi} \epsilon^{\mu \nu \rho} A_\mu \p_\nu f_\rho,
\eeq
by integrating out $A_\mu$ 
\beq
\epsilon^{\mu \nu \rho} \p_\nu f_\rho = 0 ~~~~ \Longrightarrow ~~~~ f_\mu = \p_\mu \gamma.
\eeq
On the other hand, if we integrating out $f_\mu$ as
\beq
f^{\mu}  = \frac{u}{2\pi} \epsilon^{\mu \nu \rho} \p_\nu A_{\rho} = \frac{u}{4\pi} \epsilon^{\mu \nu \rho} F_{\nu \rho}, 
\eeq
we obtain the standard Maxwell action 
\beq
\mathcal L = - \frac{1}{4g^2} F_{\mu \nu} F^{\mu \nu}, \hs{10} 
\mbox{with} ~~ g^2 = \frac{4\pi^2}{u}. 
\label{eq:S_gauge}
\eeq
Therefore, the free action of for 
the Abelian gauge field Eq.\,\eqref{eq:S_gauge} and 
the periodic scalar field Eq.\,\eqref{eq:S_scalar} are 
physically equivalent. 
We can check that the winding number of $\gamma$ corresponds to the electric charge
\beq
\oint dx_i \, \p_i \gamma = \frac{2\pi}{g^2} \oint dx_i \, \epsilon^{ij} F_{j0} = \frac{2\pi}{g^2} \int d^2 x \, \p_i F_{i0} \in 2\pi \Z.
\eeq
This implies that a charged particle and a vortex are 
exchanged by this duality transformation.  

In the presence of the BF coupling between the field strength $F_{\mu \nu}$ and 
another gauge field $B_\mu$
\beq
\mathcal L = - \frac{1}{4g^2} F_{\mu \nu} F^{\mu \nu} + \frac{1}{2\pi} \epsilon^{\mu \nu \rho} B_\mu F_{\nu \rho}, 
\eeq
the corresponding scalar action takes the form
\beq
\mathcal L = - \frac{1}{2u} (\p_\mu \gamma + B_\mu)^2. 
\eeq
Therefore, the introduction of the BF coupling corresponds to 
the gauging of the $U(1)$ symmetry.

\section{The dual pair of theories}\label{appendix:dual_pair}
\begin{figure}[h]
\begin{center}
\begin{minipage}[h]{0.1\hsize}
\includegraphics[width=15mm]{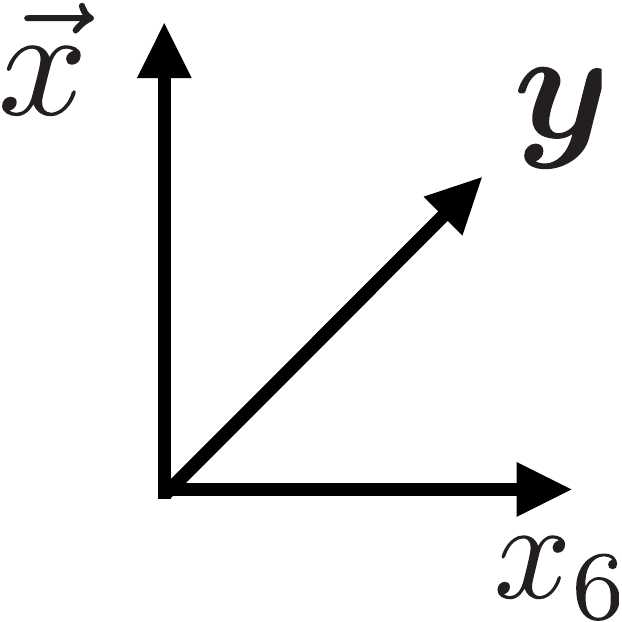}
\end{minipage}
\hs{-5}
\begin{minipage}[h]{0.45\hsize}
\begin{center}
\includegraphics[width=80mm]{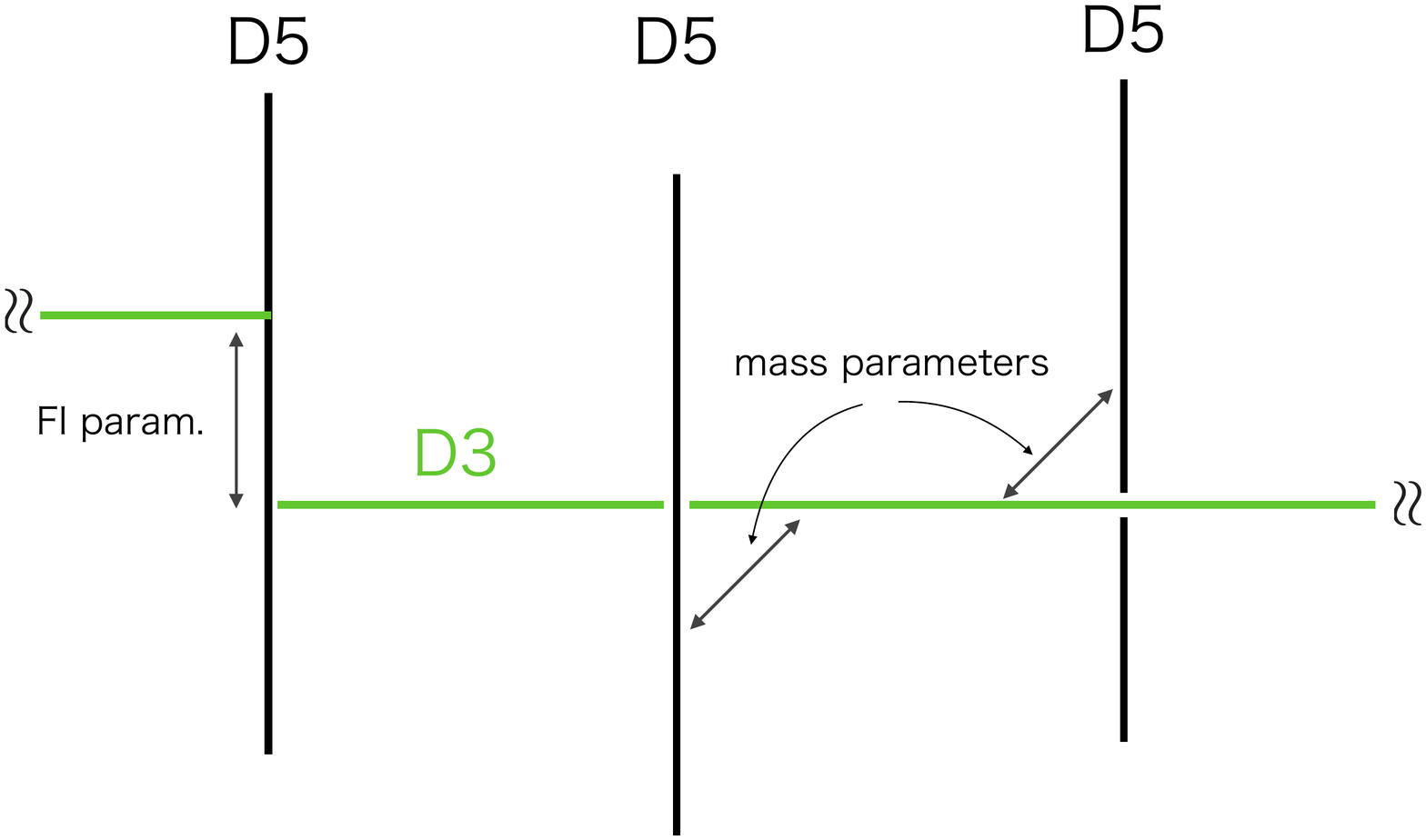} \\ \vs{-3}
(a) Theory A
\end{center}
\end{minipage}
\hs{-5}
\begin{minipage}[h]{0.45\hsize}
\begin{center}
\includegraphics[width=80mm]{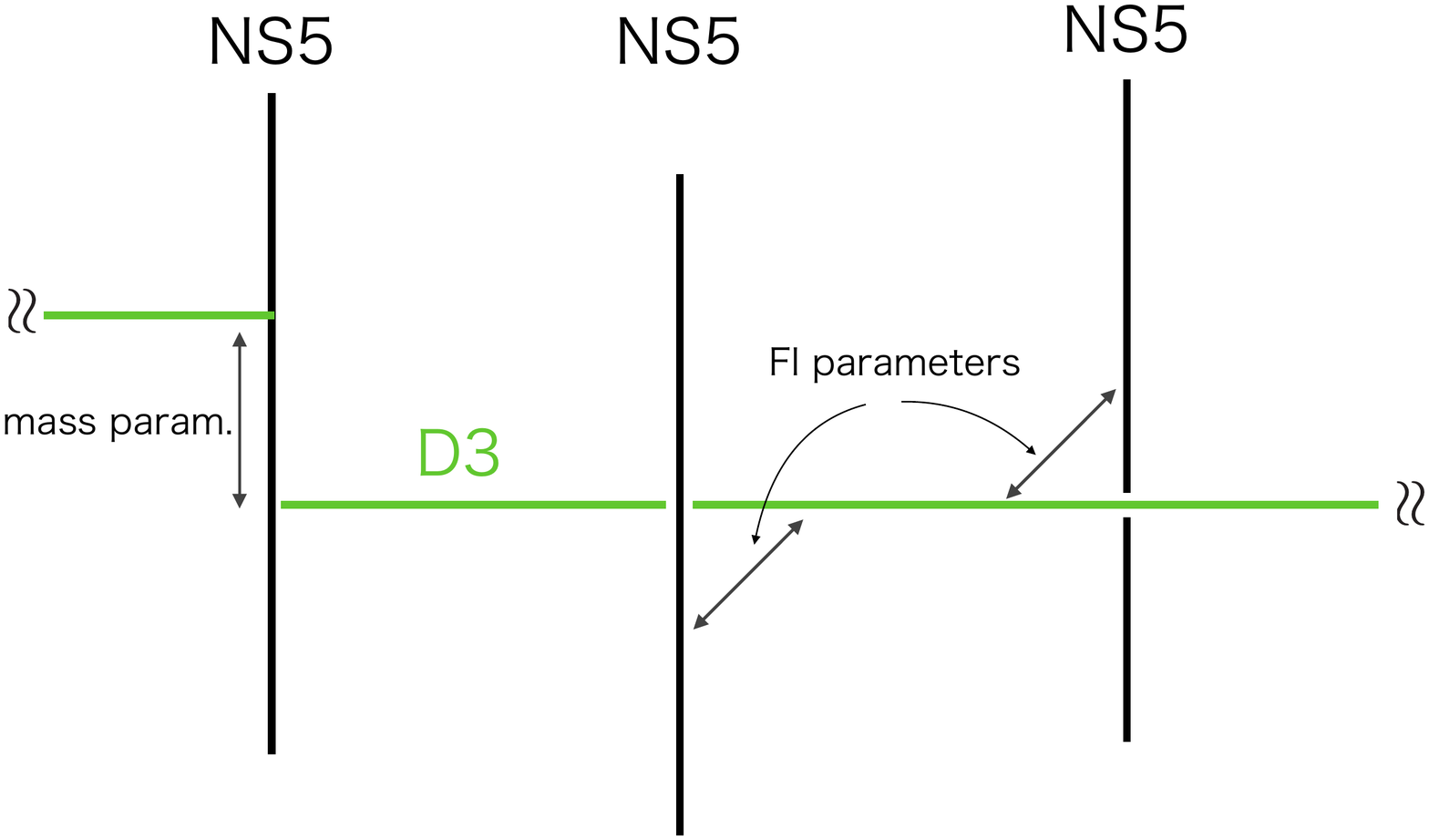} \\ \vs{-3}
(b) Theory B
\end{center}
\end{minipage} \\ \vs{5}
\begin{tabular}{c|cccccccccc}
&$x^0$&$x^1$&$x^2$&$x^3$&$x^4$&$x^5$&$x^6$&$x^7$&$x^8$&$x^9$\\
\hline
D3-brane &$\bullet$&$\bullet$&$\bullet$&
&&&$\bullet$&&& \\
\hline
$N$ 5-branes &$\bullet$&$\bullet$&$\bullet$&$\bullet$&$\bullet$
&$\bullet$&&&&
\end{tabular}
\caption{Brane configurations for the dual pair with $N=3$. 
The $x^6$ direction is compactified on $S^1$ with a twisted boundary condition 
in such a way that the position of the D3-brane is shifted along a vector in the 3d subspace $\vec x=(x_3,x_4,x_5)$. 
The shift vector corresponds to the FI parameter in Theory A and mass parameters in Theory B, 
whereas the positions of five-branes in the 3d subspace $\boldsymbol y=(x_7,x_8,x_9)$ 
are identified with mass and FI parameters in in Theory A and B, respectively.}
\label{fig:brane_config}
\end{center}
\end{figure}

In this appendix, we summarize 
the details of the three dimensional mirror symmetry
in $\mathcal N =4$ supersymmetric theories. 
We consider the following dual pairs of theories: 
\begin{itemize}
\item
Theory A: $U(1)^N$ gauge theory with hypermultiplets parameterizing $(\R^4)^N \times (\R^3 \times S^1)^{N-1}$.
\item
Theory B: $U(1)^{N}$ gauge theory with hypermultiplets parameterizing $(\R^4)^N \times (\R^3 \times S^1)$.
\end{itemize}
This dual pair of models are identified with 
the $S$-dual pair of the effective theories on the D3-branes in the Hanany-Witten type brane configurations \cite{Hanany:1996ie}
(see Fig.\,\ref{fig:brane_config}).
The details of the brane configurations are summarized below. 

The $R$-symmetry of 3d $\mathcal N =4$ supersymmetry algebra 
is $SU(2)_R \times SU(2)_L$ corresponding to . 
We use bold face symbols to denote triplet of $SU(2)_R$ and 
symbols with an arrow for triplets of $SU(2)_L$.

\subsection{Theory A}
The bosonic part of the action of Theory A takes the form
\beq
S_A = \int d^3 x \left( \sum_{i=1}^N \mathcal L_i + \sum_{a=1}^{N} \mathcal L_a \right),
\eeq
where 
\beq
\mathcal L_i &=& - \frac{1}{g_i^2} \left[ \frac{1}{2} (F_{\mu \nu}^i)^2 + (\p_\mu \boldsymbol \Sigma_i)^2 + (\vec D_i)^2 \right] - |\D_\mu H_i|^2 - (\boldsymbol \Sigma_i - \boldsymbol m_i)^2 |H_i|^2, \phantom{\Bigg[} \\
\mathcal L_a &=& - \, \frac{1}{2} \, \left[ u_a (\p_\mu \vec X_a)^2 + \frac{1}{u_a} \left( \p_\mu \chi_a + \alpha_a^i A_\mu^i \right)^2 + \frac{1}{u_a} \left( \alpha_a^i \boldsymbol \Sigma_i \right)^2 \right]. \phantom{\Bigg[}
\eeq
There are three types of multiplets in this model ($\cdots$ denotes fermionic partners): 
\begin{itemize}
\item
vector multiplets $(A_\mu^i, \boldsymbol \Sigma_i, \vec D_i, \cdots)$ : 
Each vector multiplet consists of a $U(1)$ gauge field $A_\mu^i$, 
an $SU(2)_L$ triplet scalar $\boldsymbol \Sigma_i$ and
an $SU(2)_R$ triplet auxiliary fields $\vec D_i$. 
\item
$\R^4$ hypermultiplets $(H_i,\cdots)$: 
$H_i$ is the $SU(2)_R$ doublet scalar 
in each hypermultiplet (2-component column vector)
and it is charged under the gauge field 
$A_\mu^i$ $(\D_\mu H_i \equiv (\p_\mu + i A_\mu^i) H_i)$. 
\item
$\R^3 \times S^1$ hypermultiplets $(\vec X_a, \chi_a, \cdots)$: 
The $SU(2)_L$ triplet $\vec X_a$ and the periodic scalar $\chi_a$
parametrizes $\R^3$ and $\times S^1$, respectively. 
$\chi_a$ is coupled to the gauge fields 
via the Stueckelberg type interactions with coefficients
\beq
\alpha_a^i = \delta_{a}{}^i - \delta_{a}{}^{i-1},
\eeq
where the Kronecker delta $\delta_{a}{}^{0}$ is interpreted as $\delta_{a}{}^N$. 
\end{itemize}
The auxiliary fields $\vec D_i$ are given by
\beq
\vec D_i = \frac{g_i^2}{2} \left( H_i^\dagger \vec \tau \, H_i + \alpha_a^i \vec X_a - \vec \xi \right),
\eeq
where $\vec \tau$ are the Pauli matrices. 
The parameters of this model are 
the gauge coupling constants $g_i$, 
the periods of $(S^1)^N$ $\propto u_a^{-1}$, 
the $SU(2)_L$ triplet masses $\boldsymbol m_i$ and 
the $SU(2)_R$ triplet Fayet-Iliopoulos (FI) parameter $\vec \xi$.
Note that the overall part of $(\R^3 \times S^1)^N$ parameterized 
by $\sum_{a=1}^N (\vec X_a, \chi_a)$ 
is decoupled from the other fields,
so that the interacting part of 
the Lagrangian essentially contain only $(\R^3 \times S^1)^{N-1}$. 

\subsection{Theory B}
The bosonic part of the action of Theory B takes the form
\beq
S_B ~=~ \int d^3 x \left( \sum_{A=1}^N \mathcal L_A + \mathcal L_0 \right),
\eeq
where 
\beq
\mathcal L_A &=& - \frac{1}{e_A^2} \left[ \frac{1}{2} (f_{\mu \nu}^A)^2 + (\p_\mu \vec \sigma_A)^2 + (\boldsymbol D_A)^2 \right] 
- |\D_\mu \phi_A|^2 - (\vec \sigma_A)^2 |\phi_A|^2, \phantom{\Bigg[} \\
\mathcal L_0 &=& - \, \frac{1}{2} \ \left[ u (\p_\mu \boldsymbol Y)^2 + \frac{1}{u} \left( \p_\mu \chi + \sum_{A=1}^N a_\mu^A \right)^2 + \frac{1}{u} \left( \sum_{A=1}^N \vec \sigma_A - \vec m \right)^2 \right]. 
\eeq
The field content of this model is
\begin{itemize}
\item
vectormultiplets $(a_\mu^A, \vec \sigma_A, \boldsymbol D_A,\cdots)$: 
Each vector multiplets consists of a gauge field $a_\mu^A$, 
an $SU(2)_R$ triplet scalar $\vec \sigma_A$ and
an $SU(2)_L$ triplet auxiliary field $\boldsymbol D_A$. 
\item
$\R^4$ hypermultiplets $(\phi_A,\cdots)$: 
The $SU(2)_L$ doublet scalar $\phi_A$ in each hypermultiplet
is charged under the gauge field $a_\mu^A$ $(\D_\mu \phi_A \equiv (\p_\mu + i a_\mu^A)\phi_A)$.
\item
$\R^3 \times S^1$ hypermultiplet $(\boldsymbol Y, \chi, \cdots)$: 
The $SU(2)_L$ triplet $\boldsymbol Y$ and the singlet $\chi$ 
are scalars in the hypermultiplet parameterizing $S^1 \times \R^3$. 
\end{itemize}
The auxiliary fields $\boldsymbol D_A$ are given by
\beq
\boldsymbol D_A = \frac{e_A^2}{2} \left( \phi_A^\dagger \boldsymbol \tau \, \phi_A + \boldsymbol Y - \boldsymbol \xi_A \right),
\eeq
where $\boldsymbol \tau$ are the Pauli matrices. 
$e_A$ are gauge coupling constants, $u$ is a parameter related to the period of $\chi$, 
$\boldsymbol \xi_A$ are $SU(2)_L$ triplet FI parameters and 
$\vec m$ is a $SU(2)_R$ triplet mass parameter.

\subsection{Duality}
Both theories have Coulomb and Higgs branches 
in the absence of the masses and the FI parameters. 
The 3d mirror symmetry exchanges the two branches of the dual pair
\beq
\mbox{Coulomb (Higgs) branch of Theory A} ~~ \Longleftrightarrow ~~ \mbox{Higgs (Coulomb) branch of Theory B}.  
\eeq
We can easily check the agreement of the numbers of the low-energy degrees of freedom
\beq
{\rm dim} \, \mathcal M_A^{\rm Coulomb} = {\rm dim} \, \mathcal M_B^{\rm Higgs} = 4, \hs{10}
{\rm dim} \, \mathcal M_A^{\rm Higgs} = {\rm dim} \, \mathcal M_B^{\rm Coulomb} = 4 N.
\eeq
The Higgs brach effective action can be obtained 
by the standard hyperK\"ahler quotient construction, 
whereas the Coulomb branch effective theories can be obtained 
by integrating out the charged matters, 
which gives only one-loop corrections due to the supersymmetry. 

\paragraph{Coulomb branch of Theory A = Higgs branch of Theory B \\}
If the mass parameters are turned on $\boldsymbol m_i \not = 0$ in Theory A, 
the Higgs branch is lifted and 
the low energy dynamics is described by the effective theory on the Coulomb branch moduli space
parameterized by $\boldsymbol \Sigma = \frac{1}{N} \sum_{i=1}^{N} \boldsymbol \Sigma_i$ 
and the dual photon $\gamma$ corresponding to $\sum_{i=1}^{N} A_\mu^i$. 
We can show that the moduli space metric, 
which is one-loop exact, is given by the multi-center Taub-NUT metric
\beq
ds^2 = H d \boldsymbol \Sigma \cdot d\boldsymbol \Sigma + H^{-1} (d \gamma + \boldsymbol \omega \cdot d \boldsymbol \Sigma)^2, \hs{10}
H = \frac{4\pi^2}{g^2} + \sum_{i=1}^N \frac{1}{|\boldsymbol \Sigma - \boldsymbol m_i|},
\label{eq:CoulombA}
\eeq
where $\boldsymbol \omega$ and the parameter $g$ is given by
\beq
\frac{\p}{\p \boldsymbol \Sigma} \times \boldsymbol \omega = \frac{\p}{\p \boldsymbol \Sigma} U,
\hs{10}
\frac{1}{g^2} = \frac{1}{N} \sum_{i=1}^N \frac{1}{g_i^2}. 
\eeq

On the other hand, when the FI parameters are turned on $\boldsymbol \xi_A \not =0$ in Theory B, 
the Coulomb branch is lifted and low energy physics is described by a non-linear sigma model on
the Higgs branch parameterized by $\boldsymbol Y$ and $\chi$. 
The standard hyperK\"ahler quotient procedure \cite{Gibbons:1996nt}
gives the multi-center Taub-NUT metric \eqref{eq:CoulombA} with
\beq
\boldsymbol \Sigma \rightarrow 2 \pi \boldsymbol Y, \hs{10} 
\gamma \rightarrow \chi, \hs{10}
\boldsymbol m_i \rightarrow  2\pi \boldsymbol \xi_A, \hs{10}
\frac{4\pi^2}{g^2} \rightarrow u.
\label{eq:para_map1}
\eeq

\paragraph{Higgs branch of Theory A = Coulomb branch of Theory B\\}
When the FI parameter is turned on $\vec \xi \not = 0$ in Theory A, 
the Coulomb branch is lifted and  
the low-energy effective dynamics is described 
by the Higgs branch non-linear sigma model. 
The hyperK\"ahler quotient procedure gives the metric
\beq
ds^2 = U_{ab} \, d \vec X_a \cdot d \vec X_b + (U^{-1})_{ab} (d \chi_a + \vec \Omega_{ac} \cdot d \vec X_c) (d \chi_b + \vec \Omega_{bd} \cdot d \vec X_d),
\label{eq:HiggsA}
\eeq
with
\beq
U_{ab} = u_a \delta_{ab} + \frac{1}{2} \sum_{i=1}^N \frac{\alpha_a^i \alpha_b^i}{|\alpha_a^i \vec X_a - \vec \xi|}, \hs{10}
\Omega_{ab} = \frac{1}{2} \sum_{i=1}^N \alpha_a^i \alpha_b^i \ \vec \omega_i,
\eeq
where $\vec \omega_i$ is the Dirac monopole connection
\beq
\frac{\p}{\p \vec x_i} \times \vec \omega_i = - \frac{\vec x_i \ }{|\vec x_i|^3}, \hs{10} \vec x_i = \alpha_c^i \vec X_c - \vec \xi . 
\eeq

On the other hand, if mass parameter $\vec m$ is turned on in Theory B, 
the Higgs branch is lifted and the Coulomb branch metric is given by \eqref{eq:HiggsA} 
with
\beq
\vec X_a \rightarrow \frac{1}{2\pi} \vec \sigma_A , \hs{5}
\chi_a \rightarrow \gamma_A , \hs{5}
\vec \xi \rightarrow \frac{\vec m}{2\pi}, \hs{5}
u_a \rightarrow \frac{2\pi}{e_A^2}.
\label{eq:para_map2}
\eeq

\paragraph{Discrete vacua and BPS mass spectrum \\}
For non-zero $\boldsymbol m_i$ and $\vec \xi$ in Theory A, 
there are $N$ supersymmetric vacua labeled by $j = 1, \cdots, N$: 
\beq
\vec X_a = \left\{ \begin{array}{cc} a \vec \xi & \mbox{for $a < j$} \\ 
(a-N) \vec \xi & \mbox{for $a \geq j$} \end{array} \right., \hs{10} 
\boldsymbol \Sigma_i = \boldsymbol m_j, \hs{10} 
H_i^\dagger \vec \tau \, H_i^{\phantom{\dagger}} = N \delta_{ij} \vec \xi.
\eeq
These vacua corresponds to the minima of the following potentials 
induced on the Higgs and the Coulomb moduli spaces:
\beq
V_{\rm Higgs} = (U^{-1})_{ab} \, (\alpha_a^i \boldsymbol m_i ) \cdot (\alpha_b^j \boldsymbol m_j), \hs{15}
V_{\rm Coulomb} = H^{-1} \vec \xi \cdot \vec \xi. 
\eeq
This also means that the vacua are given by the fixed points of the tri-holomorphic isometries 
acting on the Higgs and Coulomb branch moduli spaces. 
For $\vec \xi \not = 0$ and $\boldsymbol m \not = 0$ in Theory B, 
there are $N$ discrete vacua labeled by $B=1,\cdots,N$: 
\beq
\boldsymbol Y = \boldsymbol \xi_B, \hs{10}
\vec \sigma_A = \vec m \, \delta_{AB}, \hs{10}
\phi_A^\dagger \boldsymbol \tau \, \phi_A = \boldsymbol \xi_A - \boldsymbol \xi_B. 
\eeq

In the $j$-th vacuum of Theory A, 
there exist BPS vortices corresponding to 
the magnetic flux of the overall $U(1)$ gauge group. 
Correspondingly, in the $B$-th vacuum of Theory B, 
the hypermultiplet $\phi_B$ form a BPS supermultiplet. 
Their masses are given by
\beq
M_A^{\rm vortex} = 2\pi|\vec \xi| ~~~ \Longleftrightarrow ~~~
M_B^{\rm hyper} = |\vec m|.
\eeq
Similarly, 
the hypermultiplets $H^i~(i \not = j)$ in the $j$-th vacuum of Theory A
and 
the vortices in the $B$-th vacuum are exchanged under the duality transformation 

\beq
M_{A, i}^{\rm hyper} = |\boldsymbol m_i - \boldsymbol m_j| ~~~ \Longleftrightarrow ~~~
M_B^{\rm vortex} = 2\pi \left| \boldsymbol \xi_A - \boldsymbol \xi_B \right|.
\eeq

\begin{figure}[h]
\begin{minipage}[h]{0.1\hsize}
\includegraphics[width=15mm]{frame.pdf}
\end{minipage}
\begin{minipage}[h]{0.45\hsize}
\begin{center}
\includegraphics[width=60mm]{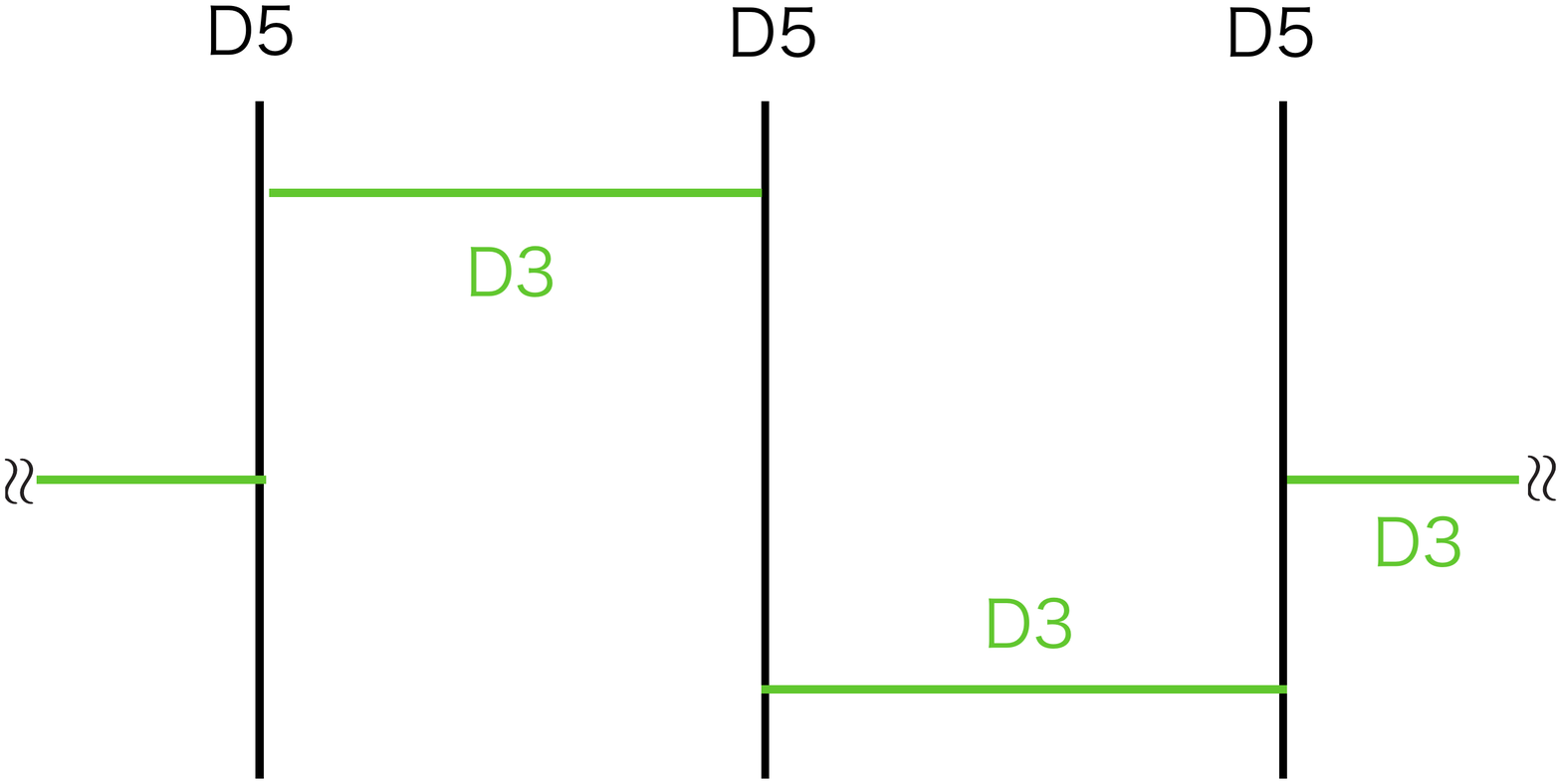} \\
(a) Higgs branch 
\end{center}
\end{minipage}
\begin{minipage}[h]{0.45\hsize}
\begin{center}
\includegraphics[width=60mm]{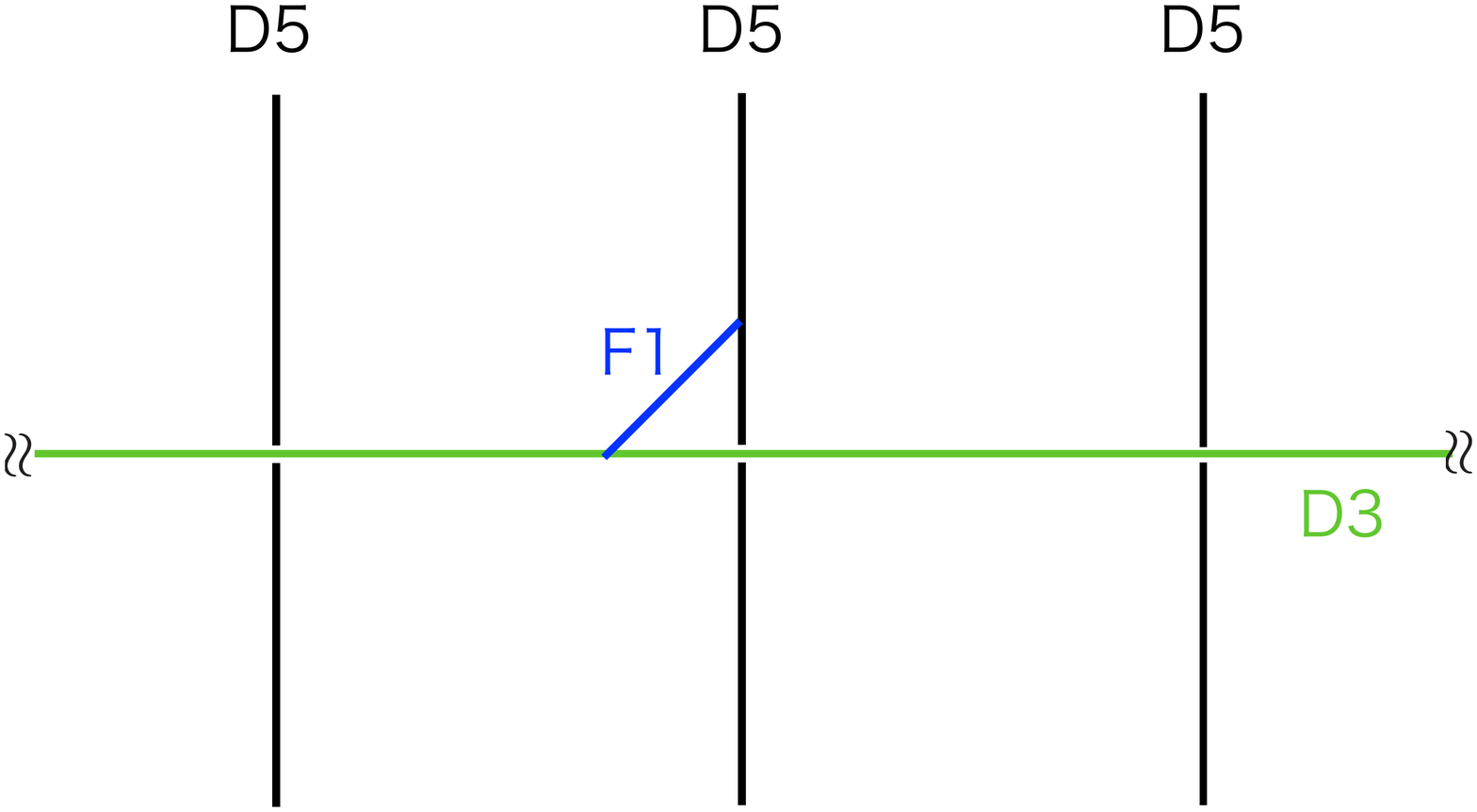} \\
(b) Coulomb branch
\end{center}
\end{minipage}
\caption{The Higgs and Coulomb branches for $\boldsymbol m_i = \vec \xi = 0$. }
\label{fig:brane_A_zero}
\end{figure}

\begin{figure}[h]
\begin{minipage}[h]{0.1\hsize}
\includegraphics[width=15mm]{frame.pdf}
\end{minipage}
\begin{minipage}[h]{0.45\hsize}
\begin{center}
\includegraphics[width=60mm]{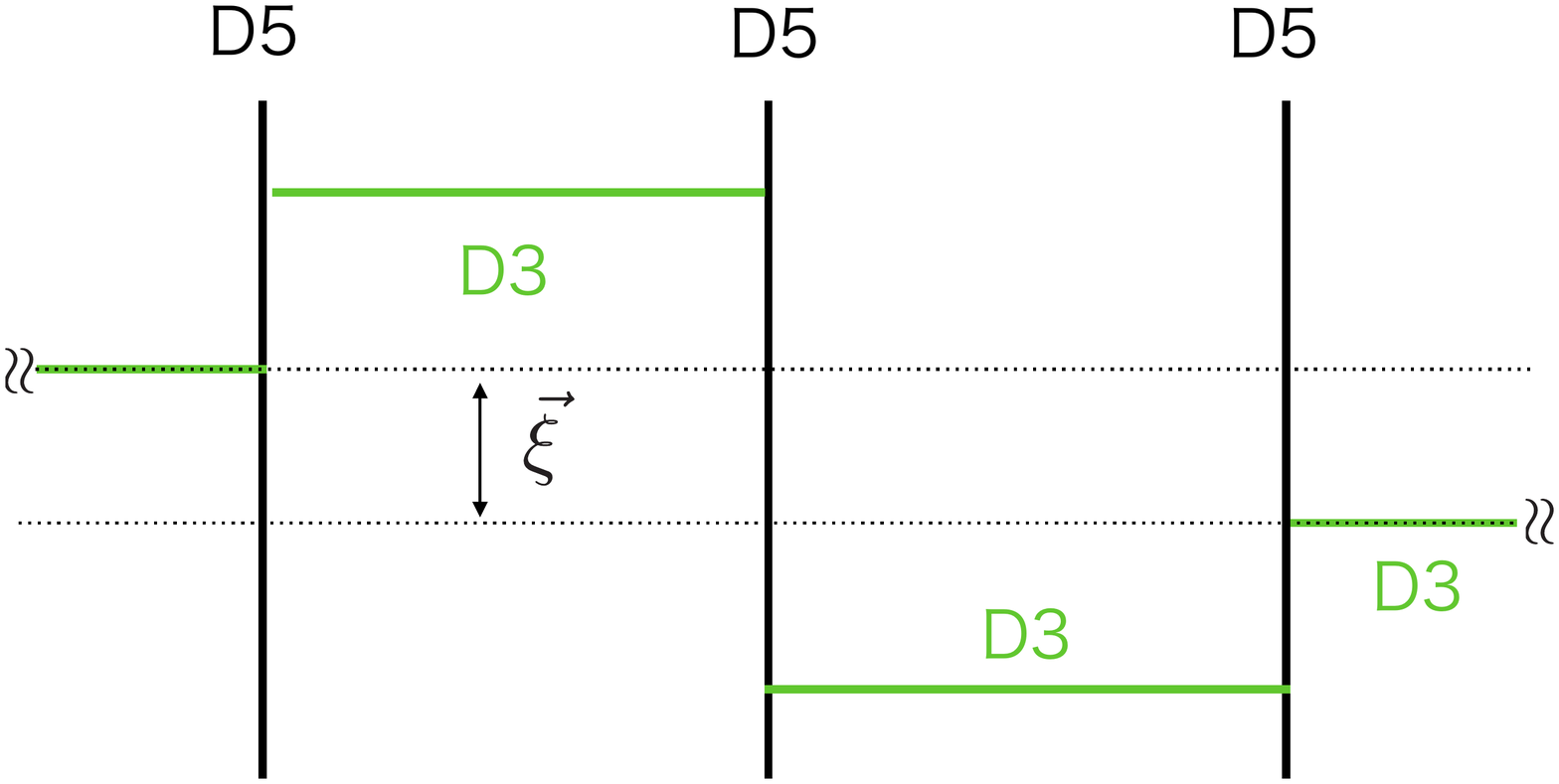} \\
(a) Higgs branch ($\vec \xi \not = 0$)
\end{center}
\end{minipage}
\begin{minipage}[h]{0.45\hsize}
\begin{center}
\includegraphics[width=60mm]{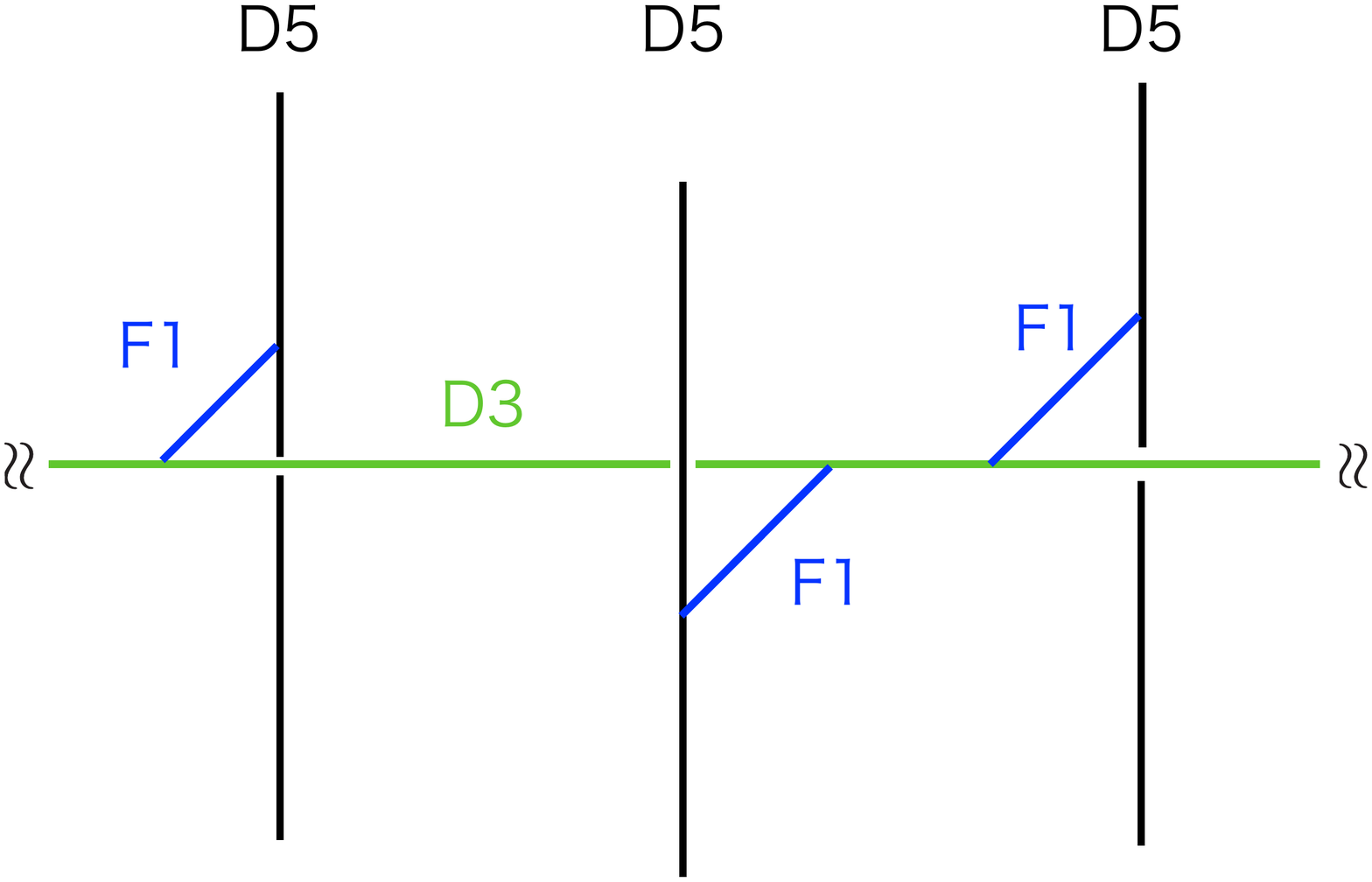} \\
(b) Coulomb branch ($\boldsymbol m_i \not = 0$)
\end{center}
\end{minipage}
\caption{The Higgs and Coulomb branches. }
\label{fig:brane_A}
\end{figure}

\paragraph{Brane construction \\}
These models can be constructed 
by using the Hanany-Witten brane configurations 
\cite{Hanany:1996ie,Witten:2009xu}. 
Figs.\,\ref{fig:brane_A_zero}-(a), \ref{fig:brane_A_zero}-(b) show 
the configuration corresponding to the Higgs and Coulomb branches of Theory A. 
The vector $\vec x$ denotes the coordinates of the $(x_7, x_8, x_9)$ directions, 
whose rotation group corresponds to the $SU(2)_R$ transformation. 
Similarly, the vector $\boldsymbol y$ denotes the $(x_3, x_4, x_5)$ directions, 
whose rotation group corresponds to the $SU(2)_L$ transformation. 
The $x_6$-direction is compactified on $S^1$ with period $l_6$. 
The scalar fields parameterizing 
the Higgs and Coulomb branches can be 
identified with the position of D3 branes: 
\beq
\vec X_a = \lim_{x_6 \rightarrow x_6^{a} + 0} \vec x_{\rm D3} ~ - \lim_{x_6 \rightarrow x_6^{a} - 0} \vec x_{\rm D3}, \hs{10}
\boldsymbol \Sigma = \boldsymbol y_{\rm D3}.
\eeq
When the FI parameter $\vec \xi$ is turned on, 
the D3 branes $\vec x_{\rm D3}$ becomes a piecewise linear function of $x_6$ 
in the supersymmetric configuration. 
We can redefine $\vec x$ so that $\vec x_{\rm D3}$ looks a piecewise constant function of $x_6$. 
Then the periodicity of $\vec x_{\rm D3}$ becomes $\vec x_{\rm D3}(x_6+l_6) = \vec x_{\rm D3}(x_6) + \vec \xi$ 
as shown in Fig.\,\ref{fig:brane_A}-(a). 
Fig.\,\ref{fig:brane_A}-(b) shows the supersymmetric state with non-zero $\boldsymbol m_i$, 
which correspond to the D5 brane positions $\boldsymbol y^{\rm D5}_i$. 

\begin{figure}[!h]
\begin{center}
\includegraphics[width=60mm]{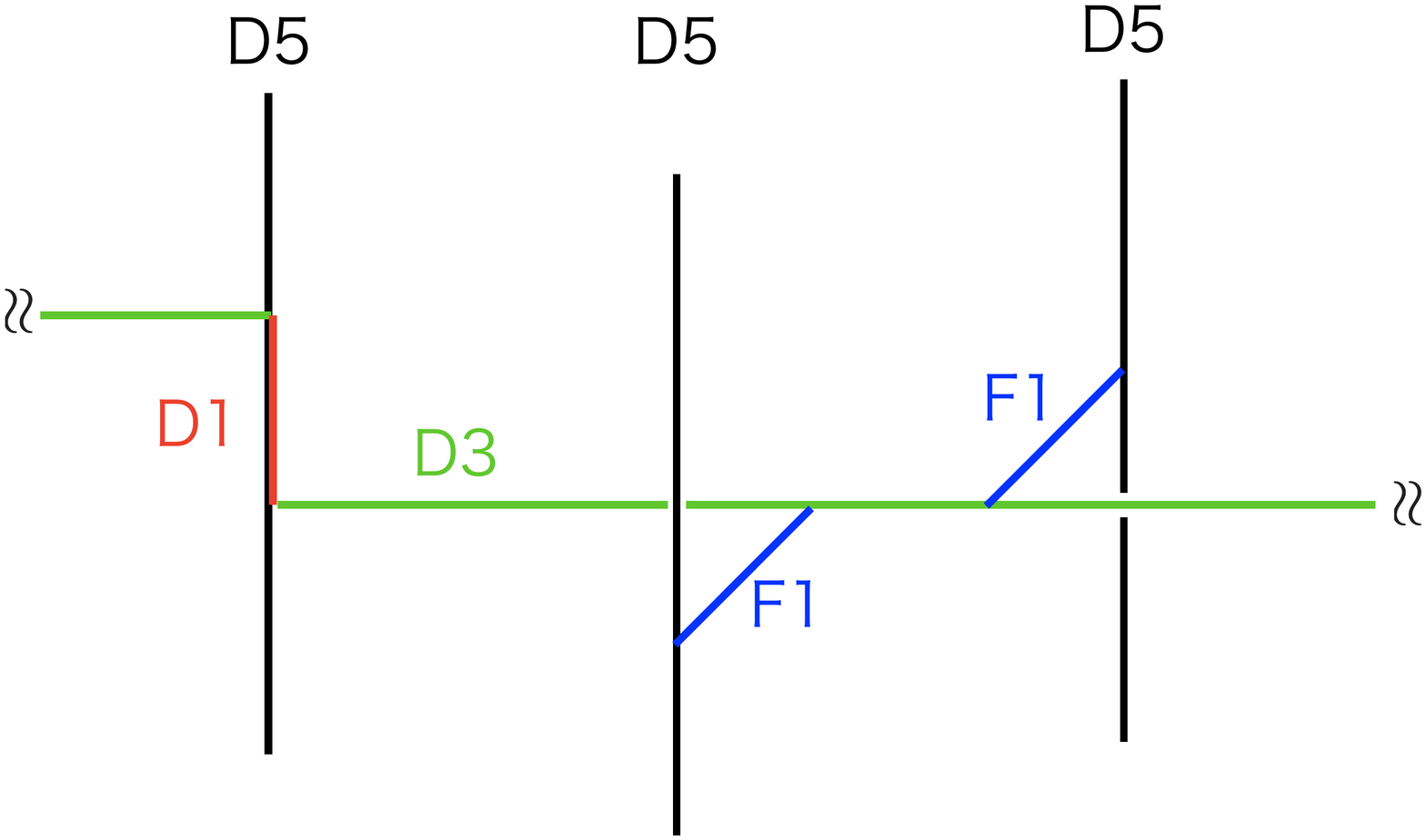} 
\end{center}
\caption{``The root of Higgs branch" }
\label{fig:RHB}
\end{figure}

Fig.\,\ref{fig:RHB} shows one of the discrete vacua in the case of $\vec \xi \not = 0$ and $\boldsymbol m_i \not = 0$. 
The D3 brane ends on one of $N$ D5 brane, so that there are $N$ supersymmetric states
corresponding to the discrete vacua of Theory A. 
D1 branes can be stretched between the end points of the D3 brane on the D5 brane, 
whereas fundamental strings can be stretched between the D3 brane and the other D5 brane. 
They can be interpreted as BPS vortices and particles with flavor charges in Theory A. 

\begin{figure}[h]
\begin{center}
\begin{minipage}[h]{0.1\hsize}
\includegraphics[width=15mm]{frame.pdf}
\end{minipage}
\hs{-5}
\begin{minipage}[h]{0.45\hsize}
\begin{center}
\includegraphics[width=60mm]{RHB.pdf} \\
(a) Theory A
\end{center}
\end{minipage}
\hs{-10}
\begin{minipage}[h]{0.45\hsize}
\begin{center}
\includegraphics[width=60mm]{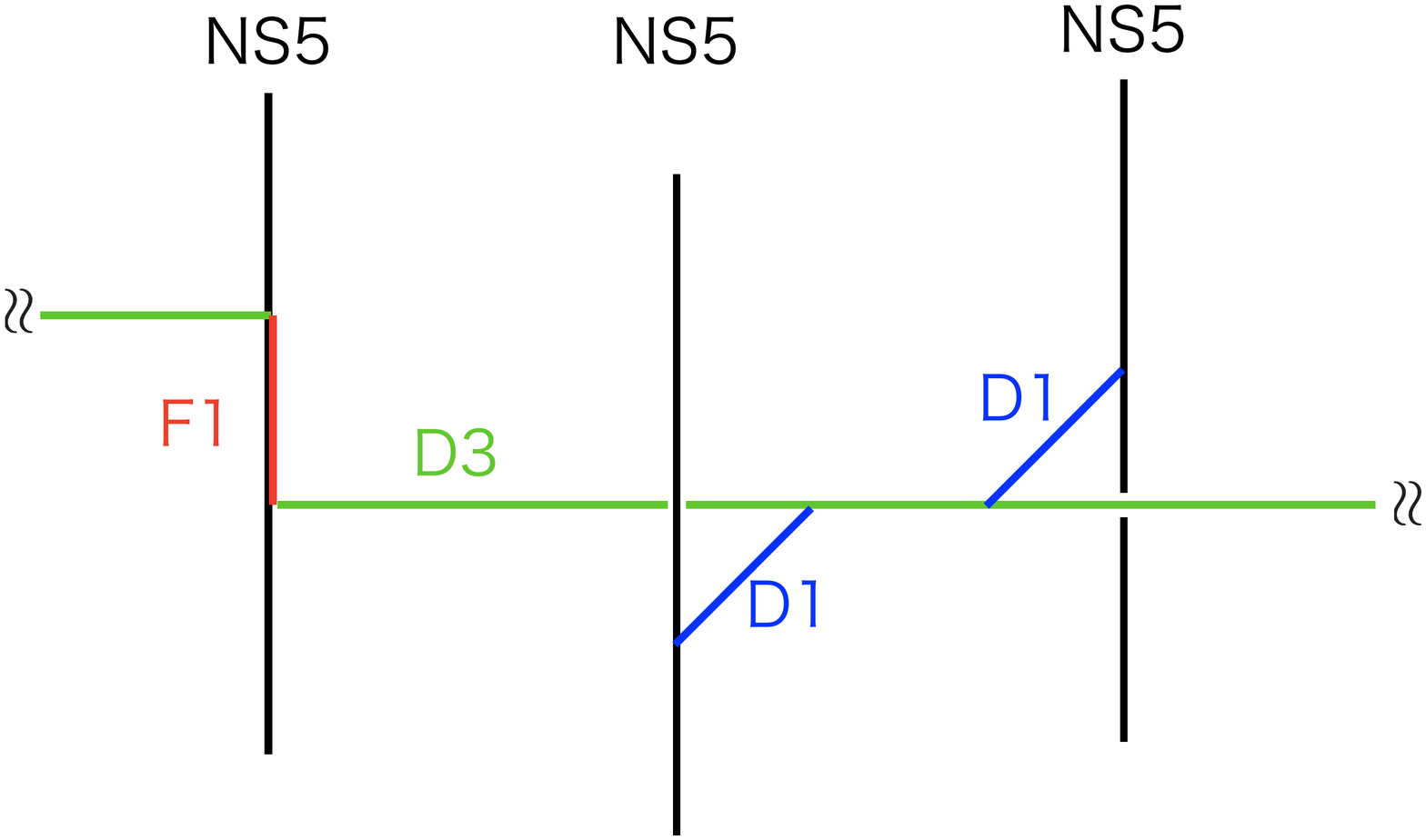} \\
(b) Theory B
\end{center}
\end{minipage}
\caption{BPS vortices (D-strings) and particles (F-strings) in Theory A and B ($N=3$). }
\label{fig:RHB0}
\end{center}
\end{figure}

The brane configuration for Theory B can be obtained 
by applying the S-duality transformation, 
under which D5 and NS5 branes and D1 and F1 strings are swapped. 
We can easily check that the Higgs and Coulomb branches are exchanged 
and the duality relation between parameters 
Eqs.\,\eqref{eq:para_map1} and \eqref{eq:para_map1} 
can be correctly read off from the dualized configuration. 
The BPS vortex and charged particle are exchanged 
by the S-duality transformation 
since they correspond to the D and F-string in Theory A and B, respectively (see Fig.\,\ref{fig:RHB0}). 
Similarly, the hypermultiplets (F-strings) in Theory A
and the vortices (D-strings) in Theory B are exchanged under the duality transformation. 

\section{BPS equations}\label{appendix:BPSeq}
In terms of the BPS equations 
\begin{align}
\mathcal B^{\pm} &= F_{12}^\pm + \frac{k}{M} D_\pm, &
\mathcal E^{\pm}_{1} &= F_{01}, &
\mathcal E^{\pm}_{2} &= F_{02} + \frac{\omega}{M} D_\pm \phantom{\Bigg[} \\
\mathcal S^{\pm}_{0} &= \p_0 \Sigma_\pm, &
\mathcal S^{\pm}_{1} &= \p_1 \Sigma_\pm, &
\mathcal S^{\pm}_{2} &= \p_2 \Sigma_\pm + \frac{m}{M} D_\pm, \phantom{\Bigg[} \\
H^{\pm}_{0} &= \D_0 \phi_\pm + i \frac{\omega}{m} \tilde \Sigma_\pm \phi_\pm, &
H^{\pm}_{1} &= \D_t \phi_\pm + i \frac{k}{m} \tilde \Sigma_\pm \phi_\pm, &
H^{\pm}_{2} &= \D_2 \phi_\pm + i \frac{M}{m} \tilde \Sigma_\pm \phi_\pm,  \phantom{\Bigg[} \\
\mathcal X_0 &= u \p_0 X, &
\mathcal X_1 &= u \p_1 X, &
\mathcal X_2 &= u \p_2 X + \frac{M}{m} \Sigma_{+-}, \phantom{\Bigg[} \\
\mathcal Y_0 &= \D_0 \chi + \frac{\omega}{m} \Sigma_{+-} , &
\mathcal Y_1 &= \D_1 \chi + \frac{k}{m} \Sigma_{+-}, &
\mathcal Y_2 &= \D_2 \chi , \phantom{\Bigg[}
\end{align}
the positive semi-definite part of the energy can be written as
\beq
\mathcal E_\pm &=& \frac{1}{g^2} \left( |\mathcal S_0^\pm|^2 + |\mathcal B^\pm|^2 + \Big| \Big| ( \mathcal E_1^\pm - i \mathcal E_2^\pm, \mathcal S_1^\pm - i \mathcal S_2^\pm ) \Big| \Big|^2_{\omega/m} \right) + \Big| \Big| ( H_1^\pm, - i H_2^\pm ) \Big| \Big|^2_{k/m}, \\
\mathcal E_0 &=& \frac{1}{2u} \left( |\mathcal X_0|^2 + |\mathcal Y_0|^2 + \Big| \Big| (\mathcal X_1 - i \mathcal X_2, i( \mathcal Y_1 - i \mathcal Y_2 )\Big| \Big|_{k/m}^2 \right)
\eeq
where 
\beq
\D_\mu \chi = \p_\mu \chi + A_\mu^+ - A_\nu^-, \hs{10} 
\tilde \Sigma_\pm = \Sigma_\pm \mp m, \hs{10}
\Sigma_{+-}  = \Sigma_+ - \Sigma_-. 
\eeq
and $||(a,b)||_\alpha^2$ is the following inner product
\beq
\big| \big| (a,b) \big| \big|_\alpha^2 ~\equiv~  \ba{cc} \bar a & \bar b \ea \ba{cc} 1 & \alpha \\ \alpha & 1 \ea \ba{c} a \\ b \ea.
\eeq


\end{document}